\newif\ifOL
\OLtrue
\documentclass[AMA,LATO2COL]{WileyNJD-v2}

\newcommand{\Papertitle}{Optimized Quantification of Spin Relaxation Times in the Hybrid State}
\articletype{Preprint - Submitted to Magnetic Resonance in Medicine}%

\received{\today}
\revised{}
\accepted{}

\usepackage{animate}
\usepackage{tikz}
\usepackage{pgfplots}
\pgfplotsset{compat=1.14,
	colormap={parula}{
		rgb255=(53,42,135)
		rgb255=(15,92,221)
		rgb255=(18,125,216)
		rgb255=(7,156,207)
		rgb255=(21,177,180)
		rgb255=(89,189,140)
		rgb255=(165,190,107)
		rgb255=(225,185,82)
		rgb255=(252,206,46)
		rgb255=(249,251,14)},
	colormap={mybluered}{
		rgb255(0cm)=(0,0,180)
		rgb255(1cm)=(0,180,180)
		rgb255(2cm)=(70,180,0)
		rgb255(3cm)=(180,180,0)
		rgb255(4cm)=(255,0,0)
		rgb255(5cm)=(128,0,0)},
	colormap={mybluewhitered_m10_2}{
		rgb255(0cm)=(0,0,255)
		rgb255(10cm)=(255,255,255)
		rgb255(12cm)=(255,0,0)},
	colormap={mybluewhitered_m1_2}{
		rgb255(0cm)=(0,0,255)
		rgb255(1cm)=(255,255,255)
		rgb255(3cm)=(255,0,0)},
	colormap={mybluewhitered_0_1}{
	rgb255(0cm)=(255,255,255)
	rgb255(1cm)=(255,0,0)},
}
\usetikzlibrary{spy,backgrounds,arrows,decorations.pathmorphing,backgrounds,positioning,fit,matrix,calc}
\pgfdeclarelayer{background}% determine background layer
\pgfdeclarelayer{foreground}% determine background layer
\pgfsetlayers{background,main,foreground}% order of layers

\RequirePackage{shellesc}
\usepgfplotslibrary{external}

\usepackage{xcolor}
\definecolor{UKLred} {RGB}{207, 25,  59}
\definecolor{UKLblue}{RGB}{ 47, 63, 157}
\definecolor{turquois}{rgb}{0,0.75,0.75}%

\usepackage{amssymb}
\usepackage{upgreek}

\usepackage[]{footmisc}

\usepackage{xr}

\newcommand{\linkcolor}{blue}
\hypersetup{
	pdfpagelayout=TwoPageLeft,
	pdfstartview=Fit,
	bookmarksopen=false,
	pdftitle={\Papertitle},
	pdfauthor={Jakob Assl\"ander},
	pdfsubject={Manuscript},
	breaklinks=true,
	colorlinks=true,
	linkcolor=\linkcolor,
	anchorcolor=black,
	citecolor=\linkcolor,
	filecolor=black,
	menucolor=red,
	urlcolor=\linkcolor,
	pdfencoding=auto
}

\begin{document}

\title{\Papertitle}
\author[1,2]{Jakob Assl\"ander*}
\author[1,2,3]{Riccardo Lattanzi}
\author[1,2,3]{Daniel K. Sodickson}
\author[1,2]{Martijn A. Cloos}
\authormark{Jakob Assl\"ander \textsc{et al}}

\address[1]{\orgdiv{Center for Biomedical Imaging, Dept. of Radiology}, \orgname{New York University School of Medicine}, \orgaddress{\state{NY}, \country{USA}}}
\address[2]{\orgdiv{Center for Advanced Imaging Innovation and Research (CAI2R), Dept. of Radiology}, \orgname{New York University School of Medicine}, \orgaddress{\state{NY}, \country{USA}}}
\address[3]{\orgdiv{Sackler Institute of Graduate Biomedical Sciences}, \orgname{New York University School of Medicine}, \orgaddress{\state{NY}, \country{USA}}}

\corres{*Jakob Assl\"ander\, Center for Biomedical Imaging, Department of Radiology, New York University School of Medicine, 650 1st Avenue, New York, NY 10016, USA.\\ \email{jakob.asslaender@nyumc.org}}

%\presentaddress{This is sample for present address text this is sample for present address text}
\fundingInfo{grants NIH/NIBIB R21 EB020096 and NIH/NIAMS R01 AR070297, and was performed under the rubric of the Center for Advanced Imaging Innovation and Research (CAI2R, www.cai2r.net), a NIBIB Biomedical Technology Resource Center (NIH P41 EB017183).}

\abstract[Summary]{
	\textbf{Purpose:}
	The analysis of optimized spin ensemble trajectories for relaxometry in the hybrid state. 
	
	\textbf{Methods:}
	First, we constructed visual representations to elucidate the differential equation that governs spin dynamics in hybrid state. Subsequently, numerical optimizations were performed to find spin ensemble trajectories that minimize the Cram\'er-Rao bound for $T_1$-encoding, $T_2$-encoding, and their weighted sum, respectively, followed by a comparison of the Cram\'er-Rao bounds obtained with our optimized spin-trajectories, as well as Look-Locker and multi-spin-echo methods. Finally, we experimentally tested our optimized spin trajectories with in vivo scans of the human brain. 
	
	\textbf{Results:}
	After a nonrecurring inversion segment on the southern hemisphere of the Bloch sphere, all optimized spin trajectories pursue repetitive loops on the northern half of the sphere in which the beginning of the first and the end of the last loop deviate from the others. The numerical results obtained in this work align well with intuitive insights gleaned directly from the governing equation. Our results suggest that hybrid-state sequences outperform traditional methods. Moreover, hybrid-state sequences that balance $T_1$- and $T_2$-encoding still result in near optimal signal-to-noise efficiency. Thus, the second parameter can be encoded at virtually no extra cost. 
	
	\textbf{Conclusion:}
	We provide insights regarding the optimal encoding processes of spin relaxation times in order to guide the design of robust and efficient pulse sequences.  We find that joint acquisitions of $T_1$ and $T_2$ in the hybrid state are substantially more efficient than sequential encoding techniques.
}

\keywords{quantitative MRI, parameter mapping, MRF,  SSFP, HSFP, optimal control}

\jnlcitation{\cname{%
\author{J. Assl\"ander}, 
\author{R. Lattanzi},
\author{D. K. Sodickson},
and 
\author{M. A. Cloos}}, 
(\cyear{2018}), 
\ctitle{\Papertitle}, \cjournal{Pre-Print}, \cvol{}.}

\maketitle

\footnotetext{Word Count: 3591}

\section{Introduction}
The dynamics of large spin-1/2 ensembles in a magnetic field are commonly described by the Bloch equations, which capture the macroscopic effects of spin-lattice and spin-spin interactions with the characteristic time constants $T_1$ and $T_2$, respectively. 
Robust and rapid quantification of these parameters is important for more objective diagnoses, longitudinal studies, and computer aided diagnosis. 
The exponential behavior of $T_1$ relaxation makes a series of inversion-recovery experiments a natural choice for quantitative $T_1$ mapping. Similarly, a series of spin-echo experiments with varying echo times provides an intuitive way to estimate $T_2$. Both methods effectively strive to trace exponential relaxation curves following a single perturbation from thermal equilibrium. However, $T_1$ and $T_2$ can also be estimated by measuring magnetization in various steady-state conditions \cite{Crawley1988,Cheng2006,Deoni2003,Deoni2004}, or even in more complex conditions that depart from the steady state\cite{Ma2013}. 
%These methods derive their time efficiency from foregoing gaps between individual measurements that are usually required for the magnetization to relax. 
Magnetic Resonance Fingerprinting (MRF)\cite{Ma2013}, for example, opened a door for flexible and efficient parameter mapping with higher signal to noise ratio (SNR) efficiency than was possible with more simplistic spin trajectories. The large number of design options for MRF pulse sequences has, however, limited optimization of MRF experiments to heuristic approaches \cite{Ma2013, Jiang2015, Cloos2016, Ma2016, Asslander2017, Jiang2017}. Furthermore, the lack of an intuitive understanding of the spin dynamics in MRF, and the over-simplified nature of the Bloch models employed in typical MRF approaches, creates a risk of introducing biases in the estimated relaxation times \cite{Asslander2017}. 
Recently, we analyzed biases introduced by $B_1$- and $B_0$-inhomogeneities, including intra-voxel dephasing, which is also known as inhomogeneous broadening, and found a sequence design space that combines the steady state's robustness with the efficiency of the transient state \cite{Asslander2018c}. During experiments designed in this space, the spin ensemble establishes a so-called \textit{hybrid state}, whose dynamics can be described by a closed-form solution of the Bloch equations. Here, we provide an in-depth analysis of hybrid-state sequences and discuss how relaxation times can be encoded efficiently with this approach. By comparing hybrid-state sequences that are optimized to encode either $T_1$ or $T_2$ alone to a joint optimization, we show that the second parameter can be measured at virtually no extra cost. 

\begin{figure*}[tbp]
	\centering
	\ifOL
	\includegraphics[]{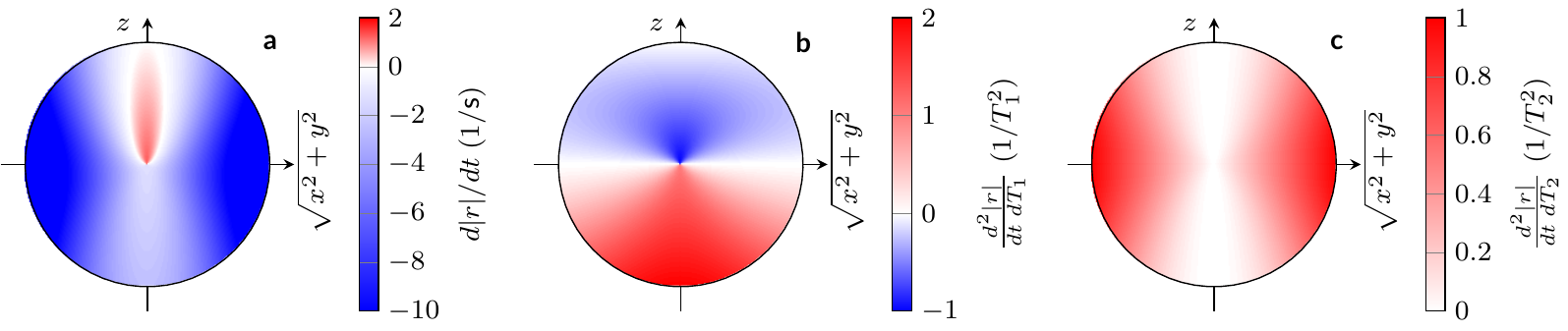}
	\else
	\begin{tikzpicture}[scale = 1]
	\def\TOne{0.781}; % wm Jiang et al. (FISP-MRF)
	\def\TTwo{0.065};
	\def\Theta{22.5};
	
	\begin{axis}[
	width=\textwidth*0.3,
	axis equal image, 
	axis lines=center, 
	xmin = -1.2, xmax = 1.2,
	ymin = -1.2, ymax = 1.2,
	xtick=\empty, ytick=\empty, ztick=\empty,
	xlabel={$\sqrt{x^2+y^2}$}, 
	ylabel={$z$}, 
	every axis x label/.style={at={(ticklabel cs:0.975)}, anchor=north,rotate=90},
	every axis y label/.style={at={(ticklabel cs:0.975)}, anchor=east},
	colormap name = mybluewhitered_m10_2, 
	colorbar,
	colorbar style={ytick={-10,-8,...,2}, ylabel=$d|r|/dt \; (1/\text{s})$, width=0.2cm, xshift=-0.1cm, yticklabel style={/pgf/number format/fixed, /pgf/number format/precision=0}},
	point meta min=-10,
	point meta max= 2,
	name=drdt,
%	at=(dzdt.right of north east),
	anchor = north west,
	clip=false,
	]
	\addplot graphics [xmin=-1,xmax=1,ymin=-1,ymax=1] {Figures/drdt.png};
	
	% Bloch-sphere
	\addplot[no marks, domain=0:360,samples=60,] ({sin(x)}, {cos(x)});
	
	% Steady-State ellipsoid 
%		\addplot[domain=0:1,samples=1000, UKLblue] ({-sqrt(\TTwo * (1/(4*\TOne) - (x - .5)^2/\TOne)}, {x});
%		\addplot[domain=0:1,samples=1000, UKLblue] ({  sqrt(\TTwo * (1/(4*\TOne) - (x - .5)^2/\TOne)}, {x});

	%	\node[above, inner sep=0mm,minimum size=5mm, rotate=90] at (axis cs:  -1.1,0) {original pSSFP};
	\node[right, inner sep=0mm,minimum size=5mm] at (axis cs:  .8,1) {\textbf{a}};
	\end{axis}
	
	\begin{axis}[
	width=\textwidth*0.3,
	axis equal image, 
	axis lines=center, 
	xmin = -1.2, xmax = 1.2,
	ymin = -1.2, ymax = 1.2,
	xtick=\empty, ytick=\empty, ztick=\empty,
	xlabel={$\sqrt{x^2+y^2}$}, 
	ylabel={$z$}, 
	every axis x label/.style={at={(ticklabel cs:0.975)}, anchor=north,rotate=90},
	every axis y label/.style={at={(ticklabel cs:0.975)}, anchor=east},
	colormap name = mybluewhitered_m1_2, 
	colorbar,
	colorbar style={ytick={-1,0,1,2}, ylabel=$\frac{d^2|r|}{dt \; dT_1} \; (1/T_1^2)$, width=0.2cm, xshift=-0.1cm, yticklabel style={/pgf/number format/fixed, /pgf/number format/precision=0}},
	point meta min=-1,
	point meta max= 2,
	name=drdtdT1,
	at=(drdt.right of north east),
	anchor = left of north west,
	xshift = 2cm,
	clip=false,
	]
	\addplot graphics [xmin=-1,xmax=1,ymin=-1,ymax=1] {Figures/drdtdT1.png};
	
	% Bloch-sphere
	\addplot[no marks, domain=0:360,samples=60,] ({sin(x)}, {cos(x)});
	
	% Steady-State ellipsoid 
	%	\addplot[domain=0:1,samples=1000, UKLblue] ({-sqrt(\TTwo * (1/(4*\TOne) - (x - .5)^2/\TOne)}, {x});
	%	\addplot[domain=0:1,samples=1000, UKLblue] ({  sqrt(\TTwo * (1/(4*\TOne) - (x - .5)^2/\TOne)}, {x});

	%	\node[above, inner sep=0mm,minimum size=5mm, rotate=90] at (axis cs:  -1.1,0) {original pSSFP};
	\node[right, inner sep=0mm,minimum size=5mm] at (axis cs:  .8,1) {\textbf{b}};
	\end{axis}
	
	\begin{axis}[
	width=\textwidth*0.3,
	axis equal image, 
	axis lines=center, 
	xmin = -1.2, xmax = 1.2,
	ymin = -1.2, ymax = 1.2,
	xtick=\empty, ytick=\empty, ztick=\empty,
	xlabel={$\sqrt{x^2+y^2}$}, 
	ylabel={$z$}, 
	every axis x label/.style={at={(ticklabel cs:0.975)}, anchor=north,rotate=90},
	every axis y label/.style={at={(ticklabel cs:0.975)}, anchor=east},
	colormap name = mybluewhitered_0_1, 
	colorbar,
	colorbar style={ylabel=$\frac{d^2|r|}{dt \; dT_2} \; (1/T_2^2)$, width=0.2cm, xshift=-0.1cm, yticklabel style={/pgf/number format/fixed, /pgf/number format/precision=1}},
	point meta min=0,
	point meta max= 1,
	name=drdtdT2,
	at=(drdtdT1.right of north east),
	anchor = left of north west,
	xshift = 2cm,
	clip=false,
	]
	\addplot graphics [xmin=-1,xmax=1,ymin=-1,ymax=1] {Figures/drdtdT2.png};
	
	% Bloch-sphere
	\addplot[no marks, domain=0:360,samples=60,] ({sin(x)}, {cos(x)});
	
	% Steady-State ellipsoid 
	%	\addplot[domain=0:1,samples=1000, UKLblue] ({-sqrt(\TTwo * (1/(4*\TOne) - (x - .5)^2/\TOne)}, {x});
	%	\addplot[domain=0:1,samples=1000, UKLblue] ({  sqrt(\TTwo * (1/(4*\TOne) - (x - .5)^2/\TOne)}, {x});

	%	\node[above, inner sep=0mm,minimum size=5mm, rotate=90] at (axis cs:  -1.1,0) {original pSSFP};
	\node[right, inner sep=0mm,minimum size=5mm] at (axis cs:  .8,1) {\textbf{c}};
	\end{axis}
	\end{tikzpicture}
	\fi
	\caption{Eq.~\eqref{eq:r} describes the spin dynamics and is visualized in (\textbf{a}). The white area indicates the steady-state ellipse which separates the area in which the magnetization grows (red) and shrinks (blue). This particular sub-figure is valid for the ratio $T_1/T_2 = 781\text{ms}/65\text{ms} \approx 12$, which are values reported for brain white matter \cite{Jiang2015}. The derivatives of Eq.~\eqref{eq:r} with respect to $T_1$ and $T_2$ are depicted in (\textbf{b}) and (\textbf{c}), respectively. These plots are normalized by the respective relaxation times and are, therefore, valid for any combination of $T_1$ and $T_2$.}
	\label{fig:diff_equations}
\end{figure*}
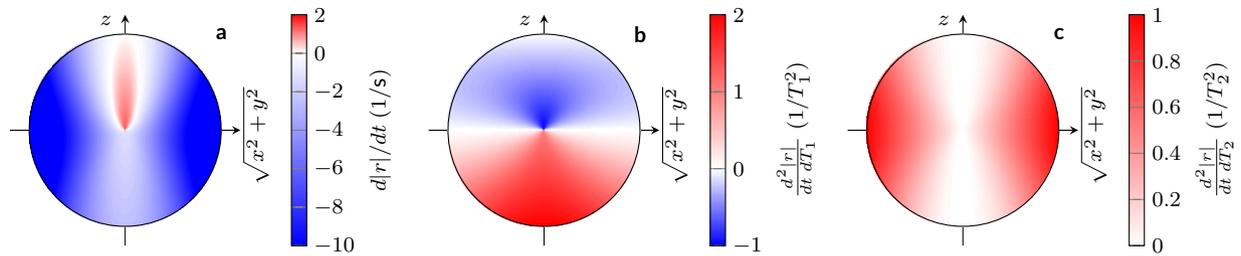

\section{Methods}
\subsection{Hybrid-State Spin Dynamics}
We start by describing the spin dynamics in spherical coordinates, which are here defined by $x = r \sin{\vartheta} \cos \varphi$, $y = r \sin{\vartheta} \sin \varphi$ and $z = r \cos{\vartheta}$, where $r$ is the radius, $\vartheta$ the polar angle or the angle between the magnetization and the $z$-axis and $\varphi$ the azimuth or the angle between the $x$-axis and the projection of the magnetization onto the $x$-$y$-plane. For practical purposes, we use the limits $-1 \leq r \leq 1$, $0 \leq \vartheta \leq \pi/2$, and $0 \leq \varphi < 2 \pi$ to uniquely identify the polar coordinates.
Recapitulating Ref. \cite{Asslander2018c}, the entire spin dynamics in hybrid state is captured by the radius $r$ of the magnetization, i.e. its magnitude combined with a sign, which is controlled only by the polar angle $\vartheta$:
\begin{equation}
	\dot{r}(t)  = - \left(\frac{\cos^2 \vartheta(t)}{T_1} +\frac{\sin^2 \vartheta(t)}{T_2} \right) r(t) + \frac{\cos \vartheta(t)}{T_1} 
	\label{eq:r}
\end{equation}
Here, $\dot{r}(t)$ denotes the derivative of $r$ with respect to time. 

%\begin{equation}
%r(t) = a(t) \cdot \left(r(0) + \frac{1}{T_1} \int_{0}^{t} \frac{\cos \vartheta(\tau)}{a(\tau)} d\tau \right)
%\label{eq:r}
%\end{equation}
%with
%\begin{equation*}
%a(\tau) = \exp \left( - \int_{0}^{\tau} \frac{\sin^2\vartheta(\xi)}{T_2} + \frac{\cos^2\vartheta(\xi)}{T_1} d\xi \right).
%\end{equation*}
%Here, $t$ denotes time and $r(0)$ the initial magnetization. 

The flip angle and the phase of the radio frequency (RF) pulses, as well as the repetition time $T_R$ have only an indirect and joint effect on the spin dynamics:
\begin{equation}
\sin^2 \vartheta = \frac{\sin^2 \frac{\alpha}{2}}{\sin^2 \frac{\phi}{2}  \cdot \cos^2 \frac{\alpha}{2} + \sin^2 \frac{\alpha}{2}},
\label{eq:vartheta}
\end{equation}
where $\phi = \omega_z T_R$ describes the phase accumulated during one repetition time and $\alpha$ denotes the flip angle.
This equation reduces to $\vartheta = \alpha/2$ for $\phi = \pi$, which we define as the on-resonance condition. In practice, $\phi = \pi$ is assigned to the on-resonant spin isochromat by the common phase increment of $\pi$ in consecutive RF pulses. 

%For the sake of completeness, the phase of the magnetization is approximated by
%\begin{equation}
%\varphi = \tan^{-1} \left( \frac{\cos \phi - E_2}{\sin \phi} \right) - \mathcal{H} \{\sin \phi \} \cdot \pi + \phi_{T_E},
%\label{eq:varphi}
%\end{equation}
%where the Heaviside function $\mathcal{H}$ disambiguates the four-quadrants and $\phi_{T_E}$ describes the phase of the magnetization accumulated between the RF pulse and the echo time $T_E$. Note that the phase of the magnetization will not be discussed in the remainder of this paper. 

\subsection{Numerical Optimizations}
Various aspects of the MRF pulse sequence design can be improved. A complete analysis considering all design parameters simultaneously is currently out of reach. Instead, this work focuses on the analysis and optimization of the $T_1$- and $T_2$-encoding power of the RF-pulse train in hybrid-state pulse sequences with balanced gradient moments, independent from the specific k-space trajectory that may be used in the imaging sequence. To this end, the Cram\'er-Rao bound \cite{Rao1945,Cramer1946} is used to provide a universal limit for the noise variance of the estimated parameters. 
Given an unbiased estimator, i.e. the fitting algorithm used to calculate the proton density ($PD$), $T_1$ and $T_2$, the noise variance of these parameters is at least as big as the corresponding Cram\'er-Rao lower bound. 
This very general and established metric has been utilized for optimizing MR parameter mapping experiments in Refs. \cite{Jones1996,Jones1997,Teixeira2017} amongst others, and for MRF in particular in Ref. \cite{Zhao2016b}. 

The hybrid state allows us to approximate a voxel's signal at the echo time $T_E = T_R/2$ by a single isochromat \cite{Asslander2018c}, similar to balanced steady-state sequences \cite{Scheffler2003}. 
With $x(t) = r(t) \sin \vartheta(t)$, we can calculate a discretized signal vector $\mathbf{x} \in \mathbb{R}^{1 \times N_t}$ from Eqs.~\eqref{eq:r} and \eqref{eq:vartheta}. This vector describes the observed signal at $N_t$ time points, and is used to calculate the Fisher information matrix $\mathbf{F} \in \mathbb{R}^{3 \times 3}$ whose entries $\mathbf{F}_{i,j}=\mathbf{b}_i^T\mathbf{b}_j/\sigma^2$ are given by 
\begin{align}
\mathbf{b}_1 &= d{\mathbf{x}}/dPD
\label{eq:Fischer_PD}\\ 
\mathbf{b}_2 &= d{\mathbf{x}}/dT_1 
\label{eq:Fischer_T1}\\ 
\mathbf{b}_3 &= d{\mathbf{x}}/dT_2.
\label{eq:Fischer_T2}
\end{align}
Here, $\sigma^2$ describes the input variance. 
The vectors $\mathbf{b}_i$ describe the derivatives of the signal evolution with respect to all considered parameters. 
By normalizing with the input variance $\sigma^2$, the duration of the experiment $T_\text{exp} = N_t T_R$, the repetition time $T_R$, and the squared relaxation time, we can define the relative Cram\'er-Rao bounds to be
\begin{align}
%rCRB(PD) &= \frac{(\mathbf{F}^{-1})_{1,1}}{\sigma^2}
%\label{eq:rCRB_PD} \\
rCRB(T_1) &= \frac{(\mathbf{F}^{-1})_{2,2}}{\sigma^2 T_1^2} \frac{T_\text{exp}}{T_R}
\label{eq:rCRB_T1} \\
rCRB(T_2) &= \frac{(\mathbf{F}^{-1})_{3,3}}{\sigma^2 T_2^2} \frac{T_\text{exp}}{T_R}.
\label{eq:rCRB_T2}
\end{align}
%In this work, we focused on the latter two, since the $PD$, as defined in this work, is modulated by the receive coil sensitivity and provides only a relative measure. 
The normalization by the variance cancels out the variance in the definition of the Fisher information matrix, the normalization by the squared relaxation time is performed in order to best reflect the $T_{1,2}$-to-noise ratio (defined as $T_{1,2}/\sigma_{T_{1,2}}$), and the normalization with $T_\text{exp}/T_R$ makes the $rCRB$ invariant to the number of measurements so that it describes the noise efficiency per unit time. In absolute numbers, the noise in the resulting parameter estimations is given by
\begin{equation}
\sigma_{T_{1,2}} \geq \sqrt{rCRB(T_{1,2}) \frac{T_R}{T_\text{exp}}} \cdot \sigma \cdot T_{1,2}.
\end{equation}
Note that the simulations published in Ref.~\cite{Asslander2018c} indicate that the actual noise in the estimated parameters is close to this bound. 

Given that $\vartheta$ is the effective drive of the spin dynamics \cite{Asslander2018c}, we optimized $\vartheta$ directly and picked the sequence parameter ($\alpha, T_R, \phi$) retrospectively (cf. Eq.~\eqref{eq:vartheta}). 
Eq.~\eqref{eq:r} is an uncoupled first order differential equation and can be solved for different boundary conditions \cite{Asslander2018c}. Here, we focus on inversion-recovery balanced hybrid-state free precession (IR-bHSFP) sequences, i.e. hybrid-state sequences that depart from thermal equilibrium by the application of an inversion pulse, which is accounted for by the boundary condition $r(0) = -1$. 

We used a Broyden-Fletcher-Goldfarb-Shanno (BFGS) algorithm \cite{DeFouquieres2011} with $rCRB(T_1)$, $rCRB(T_2)$ and $rCRB(T_1) + rCRB(T_2)$ as objective functions. The numerical optimization was based on $\vartheta(n T_R)$ with the repetition time of $T_R = 4.5~\text{ms}$ and with $n \in \{1, 2, \ldots, N_t\}$. 
The derivatives in Eqs.~\eqref{eq:Fischer_PD}-\eqref{eq:Fischer_T2} were calculated analytically, as was the gradient of the objective function with respect to each $\vartheta (n T_R)$. 

Since the $rCRB$ intrinsically compares a fingerprint to its surrounding in the parameter space, only a single set of relaxation times was used for the optimization. In particular, we used the relaxation times $T_1={781}~\text{ms}$ and $T_2={65}~\text{ms}$, which correspond to values reported for white matter\cite{Jiang2015}. All optimizations were initialized with the pattern provided in Ref. \cite{Asslander2017}. 

We performed the optimizations of IR-bHSFP sequences that exploit the full quadrant of the Bloch sphere ($0\leq{\vartheta}\leq\pi/2$), and repeated the same optimizations with $0\leq{\vartheta}\leq\pi/4$ in order to limit the flip angle to $\alpha\leq\pi/2$, ensuring consistent slice profiles by virtue of the linearity in the small tip-angle approximation \cite{Hoult1979}, and aiding compliance with safety considerations by avoiding high power large flip-angle pulses. For comparison, we also performed unconstrained optimizations of Look-Locker \cite{Look1970} and multi-spin-echo experiments.

\subsection{In Vivo Experiments}
An asymptomatic volunteer's brain was imaged following written informed consent, and according to a protocol approved by our institutional review board. Measurements were performed with the IR-bHSFP sequences that minimize $rCRB(T_1)$, $rCRB(T_2)$ and $rCRB(T_1) + rCRB(T_2)$, limited to $0 \leq {\vartheta} \leq \pi/4$. All experiments were performed on a 3T Skyra scanner (Siemens, Erlangen, Germany). The 16 head elements of the manufacturer's 20 channel head/neck coil were used for signal reception.

At the beginning of the sequence, a secant inversion pulse \cite{Silver1985} with a duration of 10.24ms was applied, followed by a spoiler gradient. The other RF-pulses were implemented as slice-selective sinc pulses with a time-bandwidth product of 2 and a duration of 1.10ms.

Spatial encoding was performed with a radial trajectory and a golden angle increment \cite{Winkelmann2007}. The spatial resolution of the maps is $1~\text{mm} \times 1~\text{mm} \times 3~\text{mm}$ at a FOV of $256~\text{mm} \times 256~\text{mm} \times 3~\text{mm}$. The readout dwell time was set to $2.4~\upmu\text{s}$ and an oversampling factor of 2 was applied. 
The total scan time of each sequence was approximately 3.8s.

The raw data were compressed to 8 virtual receive coils via SVD compression. Thereafter, image reconstruction was performed with the low rank alternating direction method of multipliers (ADMM) approach proposed in Ref. \cite{Asslander2018}. The employed dictionaries were computed with Eqs.~\eqref{eq:r} and \eqref{eq:vartheta} and covered 
%the range $T_1 (\text{s}) = 0.3 \cdot 1.02^j \; \forall \; j \in \{0, 1, \ldots, 152\}$, thus including 
the range between 300ms and 6s in logarithmic steps of 2\%. The dictionaries covered the range of $T_2$ values between 10ms and 3s in logarithmic steps of 2\%.
%, i.e. $T_2 (\text{s}) = 0.01 \cdot 1.02^j \; \forall \; j \in \{0, 1, \ldots, 289\}$. 
Slice profile correction was incorporated by adopting Ref.~\cite{Ma2017}. Off-resonance and $B_1$ correction could be applied on the basis of separately acquired maps. However, we did not correct for these effects here in order to avoid complicating the noise analysis. 
The dictionary was compressed to include singular vectors corresponding to the 6 largest singular values resulting from a singular value decomposition. The data consistency step of the ADMM alogrithm was performed with 20 conjugate gradient steps. In order to prevent non-linear effects from complicating the noise assessment, only a single ADMM iteration was performed and no spatial regularization was applied.

\section{Results}
\subsection{Visualization of the Differential Equation} \label{sec:Diff_Eq}
In order to provide some intuition for the hybrid-state spin dynamics, we visualize its governing differential equation (Eq.~\eqref{eq:r}) in Fig.~\ref{fig:diff_equations}a. One can identify the so-called steady-state ellipse \cite{Carr1958, Freeman1971, Hennig2002, Lapert2013}, which separates areas of the Bloch sphere in which the magnetization grows (red) or shrinks (blue). The absolute value of the derivative of Eq.~\eqref{eq:r} with respect to $T_1$ is in the upper limit twice as high on the southern hemisphere than on the northern hemisphere (note the asymmetric color bar in Fig.~\ref{fig:diff_equations}b), which indicates that $T_1$ can be encoded faster on the southern hemisphere. A second benefit of inversion recovery experiments becomes evident when comparing the derivative with respect to $T_1$ (b) to the one with respect to $T_2$ (c). One changes sign between the hemispheres while the other does not. Thus, encoding data on the southern and northern hemisphere minimizes the correlation between the derivatives, which should yield a superior SNR in the estimated relaxation times by virtue of a lower Cram\'er-Rao bound (cf. Eqs.~\ref{eq:Fischer_T1}-\ref{eq:Fischer_T2}).

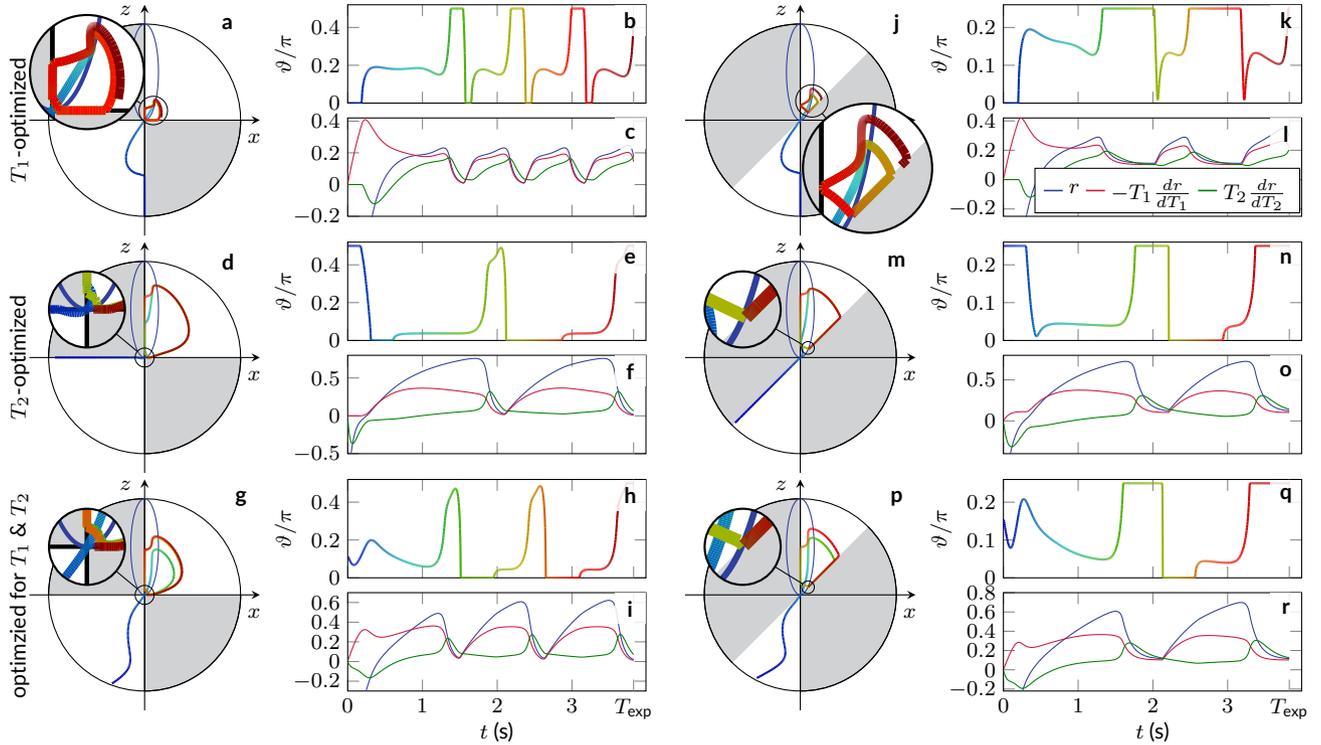
\begin{figure*}[tbp]
	\centering
	\ifOL
	\includegraphics[]{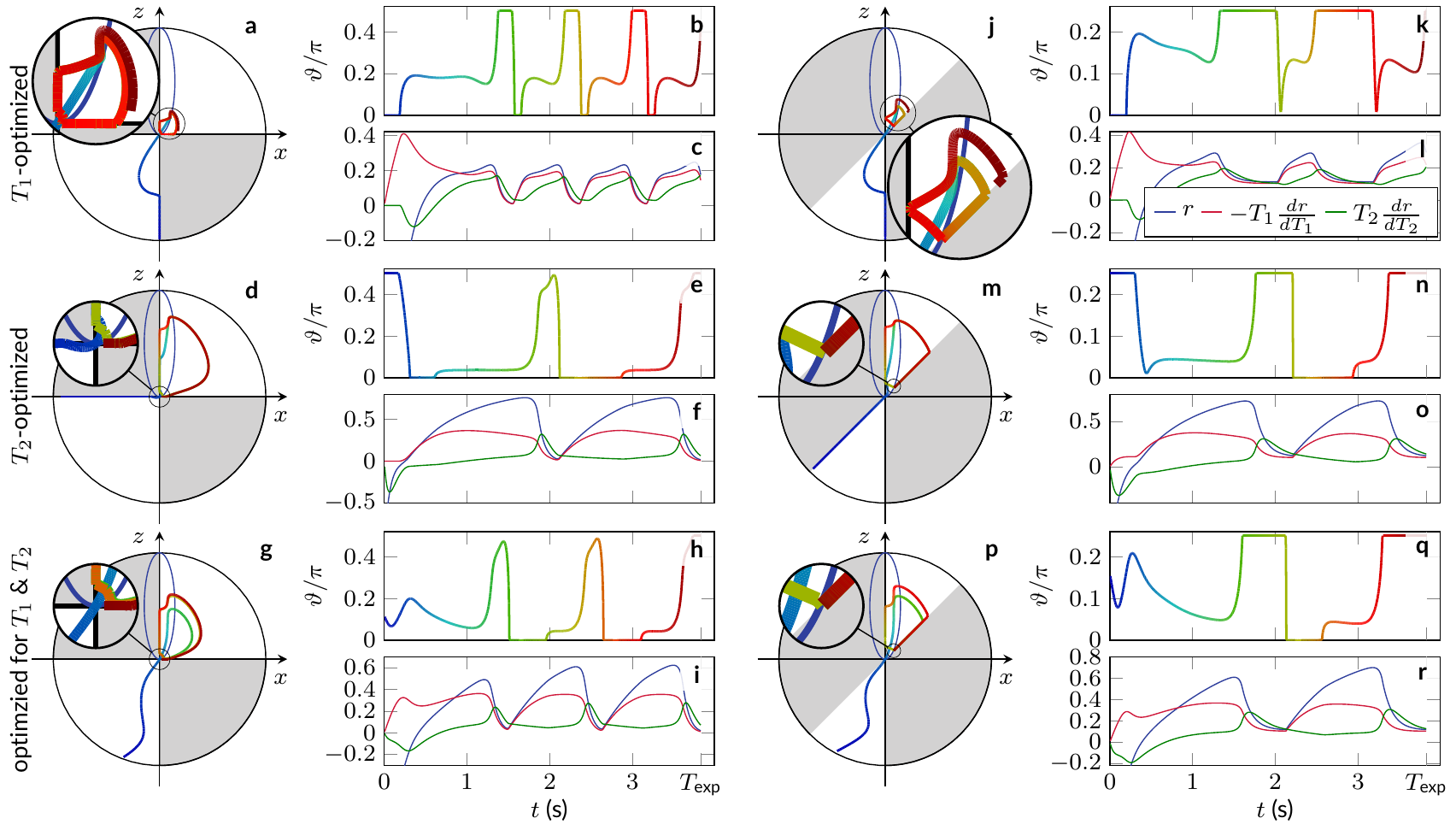}
	\else
	\begin{tikzpicture}[scale = 1]
	\def\TOne{0.781}; % wm Jiang et al. (FISP-MRF)
	\def\TTwo{0.065};
	\def\Theta{22.5};

	\begin{scope}[spy using outlines={circle, magnification=4, connect spies}]
	\begin{axis}[
	width=\textwidth*0.3,
	axis equal image, 
	axis lines=center, 
	xmin = -1.2, xmax = 1.2,
	ymin = -1.2, ymax = 1.2,
	xtick=\empty, ytick=\empty, ztick=\empty,
	xlabel={$x$}, ylabel={$z$}, 
	every axis x label/.style={at={(ticklabel cs:0.975)}, anchor=north},
	every axis y label/.style={at={(ticklabel cs:0.975)}, anchor=east},
	colormap name = mybluered,
	point meta min=0,
	point meta max= 3.825,
	name=BlochT1,
%	at=(Blochini.below south west),
	anchor=above north west,
	clip=false
	]
	
	% Bloch-sphere
	\addplot[no marks, domain=0:360,samples=60,] ({sin(x)}, {cos(x)});
	
	% Fill pies
	\begin{pgfonlayer}{background}
	\filldraw[fill=lightgray] (axis cs: 0,0) -- (axis cs: 0, 1) arc [start angle=  90, end angle= 180, radius={transformdirectionx(1)}];
	\filldraw[fill=lightgray] (axis cs: 0,0) -- (axis cs: 0,-1) arc [start angle=270, end angle=360, radius={transformdirectionx(1)}];
	\end{pgfonlayer}
	
	% Steady-State ellipsoid 
	\addplot[domain=0:1,samples=1000, UKLblue] ({-sqrt(\TTwo * (1/(4*\TOne) - (x - .5)^2/\TOne)}, {x});
	\addplot[domain=0:1,samples=1000, UKLblue] ({  sqrt(\TTwo * (1/(4*\TOne) - (x - .5)^2/\TOne)}, {x});
	
	% Magnetization
	\addplot[mesh, point meta=explicit, thick, line join=bevel]table[x=y, y=z, meta=t_s]{Figures/Bloch_sphere_T1_r0m1_pio2.txt};
	
	% Spy
	\coordinate (spypoint) at (axis cs:0.09,0.1);
	\coordinate (spyviewer) at (axis cs:-0.6,0.5);
	\spy[black,size=1.5cm] on (spypoint) in node [fill=white] at (spyviewer);

	\node[above, inner sep=0mm,minimum size=5mm, rotate=90] at (axis cs:  -1.1,0) {$T_1$-optimized};
	\node[right, inner sep=0mm] at (axis cs:  .8,1) {\textbf{a}};
	\end{axis}
	\end{scope}
	
	\begin{axis}[
	width=\textwidth*0.22,
	height=\textheight*0.055,
	scale only axis,
	xmin=0,
	xmax=3.99,
	ymin=0,
	ymax=.52,
	extra x ticks={3.825}, extra x tick labels={},
	xticklabel=\empty,
	ylabel={$\vartheta/\pi$},
	colormap name = mybluered,
	point meta min=0,
	point meta max= 3.825,
	name=thetaT1,
	at=(BlochT1.right of north east),
	anchor=left of north west,
	]
	\addplot [mesh, thick, point meta=explicit, solid]table[x=t_s, y=theta, meta=t_s]{Figures/Bloch_sphere_T1_r0m1_pio2.txt};
	
	\node[anchor=north east, fill=white, minimum size=4mm, opacity=0.85, text opacity=1] at (rel axis cs: 1,1) {\textbf{b}};
	\end{axis}	
	
	\begin{axis}[
	width=\textwidth*0.22,
	height=\textheight*0.055,
	scale only axis,
	xmin=0,
	xmax=3.99,
	ymin=-.2,
	ymax=.42,
	extra x ticks={3.825}, extra x tick labels={},
	xticklabel=\empty,
	name=FisherT1,
	at=(thetaT1.below south east),
	anchor= north east,
	]
	\addplot [color=UKLblue, solid]table[x=t_s, y=r]{Figures/Bloch_sphere_T1_r0m1_pio2.txt};
	\addplot [color=UKLred, solid]table[x=t_s, y=drdT1]{Figures/Bloch_sphere_T1_r0m1_pio2.txt};
	\addplot [color=green!50!black, solid]table[x=t_s, y=drdT2]{Figures/Bloch_sphere_T1_r0m1_pio2.txt};
	
	\node[anchor=north east, fill=white, minimum size=4mm, opacity=0.85, text opacity=1] at (rel axis cs: 1,1) {\textbf{c}};
	\end{axis}
	
	\begin{scope}[spy using outlines={circle, magnification=4, connect spies}]
	\begin{axis}[
	width=\textwidth*0.3,
	axis equal image, 
	axis lines=center, 
	xmin = -1.2, xmax = 1.2,
	ymin = -1.2, ymax = 1.2,
	xtick=\empty, ytick=\empty, ztick=\empty,
	xlabel={$x$}, ylabel={$z$}, 
	every axis x label/.style={at={(ticklabel cs:0.975)}, anchor=north},
	every axis y label/.style={at={(ticklabel cs:0.975)}, anchor=east},
	colormap name = mybluered, 
	point meta min=0,
	point meta max= 3.825,
	name=BlochT2,
	at=(BlochT1.below south west),
	anchor=above north west,
	clip=false
	]
	
	% Bloch-sphere
	\addplot[no marks, domain=0:360,samples=60,] ({sin(x)}, {cos(x)});
	
	% Fill pies
	\begin{pgfonlayer}{background}
	\filldraw[fill=lightgray] (axis cs: 0,0) -- (axis cs: 0, 1) arc [start angle=  90, end angle= 180, radius={transformdirectionx(1)}];
	\filldraw[fill=lightgray] (axis cs: 0,0) -- (axis cs: 0,-1) arc [start angle=270, end angle=360, radius={transformdirectionx(1)}];
	\end{pgfonlayer}
	
	% Steady-State ellipsoid 
	\addplot[domain=0:1,samples=1000, UKLblue] ({-sqrt(\TTwo * (1/(4*\TOne) - (x - .5)^2/\TOne)}, {x});
	\addplot[domain=0:1,samples=1000, UKLblue] ({  sqrt(\TTwo * (1/(4*\TOne) - (x - .5)^2/\TOne)}, {x});
	
	% Magnetization
	\addplot[mesh, point meta=explicit, thick, line join=bevel]table[x=y, y=z, meta=t_s]{Figures/Bloch_sphere_T2_r0m1_pio2.txt};
	
	% Spy
	\coordinate (spypoint) at (axis cs:0,0);
	\coordinate (spyviewer) at (axis cs:-0.6,0.5);
	\spy[black,size=1cm] on (spypoint) in node (myspy) [fill=white] at (spyviewer);
	
	\node[above, inner sep=0mm,minimum size=5mm, rotate=90] at (axis cs:  -1.1,0) {$T_2$-optimized};
	\node[right, inner sep=0mm] at (axis cs:  .8,1) {\textbf{d}};
	\end{axis}
	\end{scope}
	
	\begin{axis}[
	width=\textwidth*0.22,
	height=\textheight*0.055,
	scale only axis,
	xmin=0,
	xmax=3.99,
	ymin=0,
	ymax=.52,
	extra x ticks={3.825}, extra x tick labels={},
	xticklabel=\empty,
	ylabel={$\vartheta/\pi$},
	colormap name = mybluered, 
	point meta min=0,
	point meta max= 3.825,
	name=thetaT2,
	at=(BlochT2.right of north east),
	anchor=left of north west,
	]
	\addplot [mesh, thick, point meta=explicit, solid]table[x=t_s, y=theta, meta=t_s]{Figures/Bloch_sphere_T2_r0m1_pio2.txt};
	
	\node[anchor=north east, fill=white, minimum size=4mm, opacity=0.85, text opacity=1] at (rel axis cs: 1,1) {\textbf{e}};
	\end{axis}
	
	\begin{axis}[
	width=\textwidth*0.22,
	height=\textheight*0.055,
	scale only axis,
	xmin=0,
	xmax=3.99,
	ymin=-.5,
	ymax= .8,
	extra x ticks={3.825}, extra x tick labels={},
	xticklabel=\empty,
	name=FisherT2,
	at=(thetaT2.below south east),
	anchor= north east,
	]
	\addplot [color=UKLblue,            solid]table[x=t_s, y=r        ]{Figures/Bloch_sphere_T2_r0m1_pio2.txt};
	\addplot [color=UKLred,              solid]table[x=t_s, y=drdT1]{Figures/Bloch_sphere_T2_r0m1_pio2.txt};
	\addplot [color=green!50!black, solid]table[x=t_s, y=drdT2]{Figures/Bloch_sphere_T2_r0m1_pio2.txt};
	
	\node[anchor=north east, fill=white, minimum size=4mm, opacity=0.85, text opacity=1] at (rel axis cs: 1,1) {\textbf{f}};
	\end{axis}
	
	\begin{scope}[spy using outlines={circle, magnification=4, connect spies}]
	\begin{axis}[
	width=\textwidth*0.3,
	axis equal image, 
	axis lines=center, 
	xmin = -1.2, xmax = 1.2,
	ymin = -1.2, ymax = 1.2,
	xtick=\empty, ytick=\empty, ztick=\empty,
	xlabel={$x$}, ylabel={$z$}, 
	every axis x label/.style={at={(ticklabel cs:0.975)}, anchor=north},
	every axis y label/.style={at={(ticklabel cs:0.975)}, anchor=east},
	colormap name = mybluered, 
	point meta min=0,
	point meta max= 3.825,
	name=BlochT1T2,
	at=(BlochT2.below south west),
	anchor=above north west,
	clip=false
	]
	
	% Bloch-sphere
	\addplot[no marks, domain=0:360,samples=60,] ({sin(x)}, {cos(x)});
	
	% Fill pies
	\begin{pgfonlayer}{background}
	\filldraw[fill=lightgray] (axis cs: 0,0) -- (axis cs: 0, 1) arc [start angle=  90, end angle= 180, radius={transformdirectionx(1)}];
	\filldraw[fill=lightgray] (axis cs: 0,0) -- (axis cs: 0,-1) arc [start angle=270, end angle=360, radius={transformdirectionx(1)}];
	\end{pgfonlayer}
	
	% Steady-State ellipsoid 
	\addplot[domain=0:1,samples=1000, UKLblue] ({-sqrt(\TTwo * (1/(4*\TOne) - (x - .5)^2/\TOne)}, {x});
	\addplot[domain=0:1,samples=1000, UKLblue] ({  sqrt(\TTwo * (1/(4*\TOne) - (x - .5)^2/\TOne)}, {x});
	
	% Magnetization
	\addplot[mesh, point meta=explicit, thick, line join=bevel]table[x=y, y=z, meta=t_s]{Figures/Bloch_sphere_T1T2_r0m1_pio2.txt};
	
	% Spy
	\coordinate (spypoint) at (axis cs:0,0);
	\coordinate (spyviewer) at (axis cs:-0.6,0.5);
	\spy[black,size=1cm] on (spypoint) in node (myspy) [fill=white] at (spyviewer);
	
	\node[above, inner sep=0mm,minimum size=5mm, rotate=90] at (axis cs:  -1.1,0) {optimzied for $T_1$ \& $T_2$};
	\node[right, inner sep=0mm,minimum size=5mm] at (axis cs:  .8,1) {\textbf{g}};
	\end{axis}
	\end{scope}
	
	\begin{axis}[
	width=\textwidth*0.22,
	height=\textheight*0.055,
	scale only axis,
	xmin=0,
	xmax=3.99,
	ymin=0,
	ymax=.52,
	extra x ticks={3.825}, extra x tick labels={},
	xticklabel=\empty,
	ylabel={$\vartheta/\pi$},
	colormap name = mybluered, 
	point meta min=0,
	point meta max= 3.825,
	name=thetaT1T2,
	at=(BlochT1T2.right of north east),
	anchor=left of north west,
	]
	%	\addplot [color=UKLblue, solid]table[x=t_s, y=theta_ini]{Figures/Bloch_sphere_T1T2_r0m1_pio2.txt};
	%	\addplot [color=UKLred,   solid]table[x=t_s, y=theta     ]{Figures/Bloch_sphere_T1T2_r0m1_pio2.txt};
	\addplot [mesh, thick, point meta=explicit, solid]table[x=t_s, y=theta, meta=t_s]{Figures/Bloch_sphere_T1T2_r0m1_pio2.txt};
	
	\node[anchor=north east, fill=white, minimum size=4mm, opacity=0.85, text opacity=1] at (rel axis cs: 1,1) {\textbf{h}};
	\end{axis}	
	
	\begin{axis}[
	width=\textwidth*0.22,
	height=\textheight*0.055,
	scale only axis,
	xmin=0,
	xmax=3.99,
	ymin=-.3,
	ymax=.7,
	extra x ticks={3.825}, extra x tick labels={$T_\text{exp}$},
	xlabel={$t~\text{(s)}$},
	xlabel style={yshift=0.15cm},
	name=FisherT1T2,
	at=(thetaT1T2.below south east),
	anchor= north east,
	]
	\addplot [color=UKLblue,            solid]table[x=t_s, y=r        ]{Figures/Bloch_sphere_T1T2_r0m1_pio2.txt};
	\addplot [color=UKLred,              solid]table[x=t_s, y=drdT1]{Figures/Bloch_sphere_T1T2_r0m1_pio2.txt};
	\addplot [color=green!50!black, solid]table[x=t_s, y=drdT2]{Figures/Bloch_sphere_T1T2_r0m1_pio2.txt};
	
	\node[anchor=north east, fill=white, minimum size=4mm, opacity=0.85, text opacity=1] at (rel axis cs: 1,1) {\textbf{i}};
	\end{axis}

	\begin{scope}[spy using outlines={circle, magnification=4, size=0.5cm, connect spies}]
	\begin{axis}[
	width=\textwidth*0.3,
	axis equal image, 
	axis lines=center, 
	xmin = -1.2, xmax = 1.2,
	ymin = -1.2, ymax = 1.2,
	xtick=\empty, ytick=\empty, ztick=\empty,
	xlabel={$x$}, ylabel={$z$}, 
	every axis x label/.style={at={(ticklabel cs:0.975)}, anchor=north},
	every axis y label/.style={at={(ticklabel cs:0.975)}, anchor=east},
	colormap name = mybluered, 
	point meta min=0,
	point meta max= 3.825,
	name=BlochT1_pio4,
	at=(thetaT1.right of north east),
	anchor=left of north west,
	xshift = 0.5cm,
	clip=false,
	]
	
	% Fill pies
	\begin{pgfonlayer}{background}
	\filldraw[fill=lightgray] (axis cs: 0,0) -- (axis cs: 0, 1) arc [start angle=  90, end angle= 225, radius={transformdirectionx(1)}];
	\filldraw[fill=lightgray] (axis cs: 0,0) -- (axis cs: 0,-1) arc [start angle=270, end angle=405, radius={transformdirectionx(1)}];
	\end{pgfonlayer}
	
	% Bloch-sphere
	\addplot[no marks, domain=0:360,samples=60,] ({sin(x)}, {cos(x)});
	
	% Steady-State ellipsoid 
	\addplot[domain=0:1,samples=1000, UKLblue] ({-sqrt(\TTwo * (1/(4*\TOne) - (x - .5)^2/\TOne)}, {x});
	\addplot[domain=0:1,samples=1000, UKLblue] ({  sqrt(\TTwo * (1/(4*\TOne) - (x - .5)^2/\TOne)}, {x});
	
	% Magnetization
	\addplot[mesh, point meta=explicit, thick]table[x=y, y=z, meta=t_s]{Figures/Bloch_sphere_T1_r0m1_pio4.txt};

	\coordinate (spypoint) at (axis cs:0.12,0.2);
	\coordinate (spyviewer) at (axis cs:0.7,-0.5);
	\spy[black,size=1.7cm] on (spypoint) in node (myspy) [fill=white] at (spyviewer);
	
%	\node[above, inner sep=0mm,minimum size=5mm, rotate=90] at (axis cs:  -1.1,0) {$T_1$-optimized};
	\node[right, inner sep=0mm,minimum size=5mm] at (axis cs:  .8,1) {\textbf{j}};
	\end{axis}
	\end{scope}
	
	\begin{axis}[
	width=\textwidth*0.22,
	height=\textheight*0.055,
	scale only axis,
	xmin=0,
	xmax=3.99,
	ymin=0,
	ymax=.26,
	extra x ticks={3.825}, extra x tick labels={},
	xticklabel=\empty,
	ylabel={$\vartheta/\pi$},
	colormap name = mybluered, 
	point meta min=0,
	point meta max= 3.825,
	name=thetaT1_pio4,
	at=(BlochT1_pio4.right of north east),
	anchor=left of north west,
	]
	\addplot [mesh, thick, point meta=explicit, solid]table[x=t_s, y=theta, meta=t_s]{Figures/Bloch_sphere_T1_r0m1_pio4.txt};
	
	\node[anchor=north east, fill=white, minimum size=4mm, opacity=0.85, text opacity=1] at (rel axis cs: 1,1) {\textbf{k}};
	\end{axis}	
	
	\begin{axis}[
	width=\textwidth*0.22,
	height=\textheight*0.055,
	scale only axis,
	xmin=0,
	xmax=3.99,
	ymin=-.25,
	ymax=.42,
	extra x ticks={3.825}, extra x tick labels={},
	xticklabel=\empty,
	name=FisherT1_pio4,
	at=(thetaT1_pio4.below south east),
	anchor= north east,
	legend entries = {$r$, $-T_1\frac{dr}{dT_1}$, $T_2\frac{dr}{dT_2}$},
	legend pos = south east,
	legend columns=3,
	legend style={
		xshift = 0.1cm,
		legend image code/.code={\draw[##1,line width=0.6pt] plot coordinates {(0cm,0cm) (0.25cm,0cm)};}
	},
	]
	\addplot [color=UKLblue,            solid]table[x=t_s, y=r        ]{Figures/Bloch_sphere_T1_r0m1_pio4.txt};
	\addplot [color=UKLred,              solid]table[x=t_s, y=drdT1]{Figures/Bloch_sphere_T1_r0m1_pio4.txt};
	\addplot [color=green!50!black, solid]table[x=t_s, y=drdT2]{Figures/Bloch_sphere_T1_r0m1_pio4.txt};
	
	\node[anchor=north east, fill=white, minimum size=4mm, opacity=0.85, text opacity=1] at (rel axis cs: 1,1) {\textbf{l}};	
	\end{axis}
	
	\begin{scope}[spy using outlines={circle, magnification=6, connect spies}]
	\begin{axis}[
	width=\textwidth*0.3,
	axis equal image, 
	axis lines=center, 
	xmin = -1.2, xmax = 1.2,
	ymin = -1.2, ymax = 1.2,
	xtick=\empty, ytick=\empty, ztick=\empty,
	xlabel={$x$}, ylabel={$z$}, 
	every axis x label/.style={at={(ticklabel cs:0.975)}, anchor=north},
	every axis y label/.style={at={(ticklabel cs:0.975)}, anchor=east},
	colormap name = mybluered, 
	point meta min=0,
	point meta max= 3.825,
	name=BlochT2_pio4,
	at=(BlochT1_pio4.below south west),
	anchor=above north west,
	clip=false,
	]
	
	% Fill pies
	\begin{pgfonlayer}{background}
	\filldraw[fill=lightgray] (axis cs: 0,0) -- (axis cs: 0, 1) arc [start angle=  90, end angle= 225, radius={transformdirectionx(1)}];
	\filldraw[fill=lightgray] (axis cs: 0,0) -- (axis cs: 0,-1) arc [start angle=270, end angle=405, radius={transformdirectionx(1)}];
	\end{pgfonlayer}
	
	% Bloch-sphere
	\addplot[no marks, domain=0:360,samples=60,] ({sin(x)}, {cos(x)});
	
	% Steady-State ellipsoid 
	\addplot[domain=0:1,samples=1000, UKLblue] ({-sqrt(\TTwo * (1/(4*\TOne) - (x - .5)^2/\TOne)}, {x});
	\addplot[domain=0:1,samples=1000, UKLblue] ({  sqrt(\TTwo * (1/(4*\TOne) - (x - .5)^2/\TOne)}, {x});
	
	% Magnetization
	\addplot[mesh, point meta=explicit, thick, line join=bevel]table[x=y, y=z, meta=t_s]{Figures/Bloch_sphere_T2_r0m1_pio4.txt};
	
	% Spy
	\coordinate (spypoint) at (axis cs:0.08,0.1);
	\coordinate (spyviewer) at (axis cs:-0.6,0.5);
	\spy[black,size=1cm] on (spypoint) in node (myspy) [fill=white] at (spyviewer);
	
%	\node[above, inner sep=0mm,minimum size=5mm, rotate=90] at (axis cs:  -1.1,0) {$T_2$-optimized};
	\node[right, inner sep=0mm,minimum size=5mm] at (axis cs:  .8,1) {\textbf{m}};
	\end{axis}
	\end{scope}
	
	\begin{axis}[
	width=\textwidth*0.22,
	height=\textheight*0.055,
	scale only axis,
	xmin=0,
	xmax=3.99,
	ymin=0,
	ymax=.26,
	extra x ticks={3.825}, extra x tick labels={},
	xticklabel=\empty,
	ylabel={$\vartheta/\pi$},
	colormap name = mybluered, 
	point meta min=0,
	point meta max= 3.825,
	name=thetaT2_pio4,
	at=(BlochT2_pio4.right of north east),
	anchor=left of north west,
	]
	\addplot [mesh, thick, point meta=explicit, solid]table[x=t_s, y=theta, meta=t_s]{Figures/Bloch_sphere_T2_r0m1_pio4.txt};
	
	\node[anchor=north east, fill=white, minimum size=4mm, opacity=0.85, text opacity=1] at (rel axis cs: 1,1) {\textbf{n}};
	\end{axis}	
	
	\begin{axis}[
	width=\textwidth*0.22,
	height=\textheight*0.055,
	scale only axis,
	xmin=0,
	xmax=3.99,
	ymin=-.4,
	ymax= .8,
	extra x ticks={3.825}, extra x tick labels={},
	xticklabel=\empty,
	%	legend entries = {$x$, $-T_1\frac{dx}{dT_1}$, $T_2\frac{dx}{dT_2}$},
	%	legend columns=1,
	%	legend pos = south east,
	%	legend style = {xshift = - 0.25cm},
	name=FisherT2_pio4,
	at=(thetaT2_pio4.below south east),
	anchor= north east,
	]
	\addplot [color=UKLblue,            solid]table[x=t_s, y=r        ]{Figures/Bloch_sphere_T2_r0m1_pio4.txt};
	\addplot [color=UKLred,              solid]table[x=t_s, y=drdT1]{Figures/Bloch_sphere_T2_r0m1_pio4.txt};
	\addplot [color=green!50!black, solid]table[x=t_s, y=drdT2]{Figures/Bloch_sphere_T2_r0m1_pio4.txt};
	
	\node[anchor=north east, fill=white, minimum size=4mm, opacity=0.85, text opacity=1] at (rel axis cs: 1,1) {\textbf{o}};
	\end{axis}
	
	\begin{scope}[spy using outlines={circle, magnification=6, connect spies}]
	\begin{axis}[
	width=\textwidth*0.3,
	axis equal image, 
	axis lines=center, 
	xmin = -1.2, xmax = 1.2,
	ymin = -1.2, ymax = 1.2,
	xtick=\empty, ytick=\empty, ztick=\empty,
	xlabel={$x$}, ylabel={$z$}, 
	every axis x label/.style={at={(ticklabel cs:0.975)}, anchor=north},
	every axis y label/.style={at={(ticklabel cs:0.975)}, anchor=east},
	colormap name = mybluered, 
	point meta min=0,
	point meta max= 3.825,
	name=BlochT1T2_pio4,
	at=(BlochT2_pio4.below south west),
	anchor=above north west,
	clip=false,
	]
	
	% Fill pies
	\begin{pgfonlayer}{background}
	\filldraw[fill=lightgray] (axis cs: 0,0) -- (axis cs: 0, 1) arc [start angle=  90, end angle= 225, radius={transformdirectionx(1)}];
	\filldraw[fill=lightgray] (axis cs: 0,0) -- (axis cs: 0,-1) arc [start angle=270, end angle=405, radius={transformdirectionx(1)}];
	\end{pgfonlayer}
	
	% Bloch-sphere
	\addplot[no marks, domain=0:360,samples=60,] ({sin(x)}, {cos(x)});
	
	% Steady-State ellipsoid 
	\addplot[domain=0:1,samples=1000, UKLblue] ({-sqrt(\TTwo * (1/(4*\TOne) - (x - .5)^2/\TOne)}, {x});
	\addplot[domain=0:1,samples=1000, UKLblue] ({  sqrt(\TTwo * (1/(4*\TOne) - (x - .5)^2/\TOne)}, {x});
	
	% Magnetization
	\addplot[mesh, point meta=explicit, thick, line join=bevel]table[x=y, y=z, meta=t_s]{Figures/Bloch_sphere_T1T2_r0m1_pio4.txt};
	
	% Spy
	\coordinate (spypoint) at (axis cs:0.08,0.08);
	\coordinate (spyviewer) at (axis cs:-0.6,0.5);
	\spy[black,size=1cm] on (spypoint) in node (myspy) [fill=white] at (spyviewer);
	
%	\node[above, inner sep=0mm,minimum size=5mm, rotate=90] at (axis cs:  -1.1,0) {optimzied for $T_1$ \& $T_2$};
	\node[right, inner sep=0mm,minimum size=5mm] at (axis cs:  .8,1) {\textbf{p}};
	\end{axis}
	\end{scope}
	
	\begin{axis}[
	width=\textwidth*0.22,
	height=\textheight*0.055,
	scale only axis,
	xmin=0,
	xmax=3.99,
	ymin=0,
	ymax=.26,
	extra x ticks={3.825}, extra x tick labels={},
	xticklabel=\empty,
	ylabel={$\vartheta/\pi$},
	colormap name = mybluered, 
	point meta min=0,
	point meta max= 3.825,
	name=thetaT1T2_pio4,
	at=(BlochT1T2_pio4.right of north east),
	anchor=left of north west,
	]
	%\addplot [color=UKLblue, solid]table[x=t_s, y=theta_ini]{Figures/Bloch_sphere_T1T2_r0m1_pio4.txt};
	%\addplot [color=UKLred, solid]table[x=t_s, y=theta]{Figures/Bloch_sphere_T1T2_r0m1_pio4.txt};
	\addplot [mesh, thick, point meta=explicit, solid]table[x=t_s, y=theta, meta=t_s]{Figures/Bloch_sphere_T1T2_r0m1_pio4.txt};
	
	\node[anchor=north east, fill=white, minimum size=4mm, opacity=0.85, text opacity=1] at (rel axis cs: 1,1) {\textbf{q}};
	\end{axis}	
	
	\begin{axis}[
	width=\textwidth*0.22,
	height=\textheight*0.055,
	scale only axis,
	xmin=0,
	xmax=3.99,
	ymin=-.22,
	ymax=.8,
	extra x ticks={3.825}, extra x tick labels={$T_\text{exp}$},
	xlabel={$t~\text{(s)}$},
	xlabel style={yshift=0.15cm},
	name=FisherT1T2_pio4,
	at=(thetaT1T2_pio4.below south east),
	anchor= north east,
	]
	\addplot [color=UKLblue,            solid]table[x=t_s, y=r        ]{Figures/Bloch_sphere_T1T2_r0m1_pio4.txt};
	\addplot [color=UKLred,              solid]table[x=t_s, y=drdT1]{Figures/Bloch_sphere_T1T2_r0m1_pio4.txt};
	\addplot [color=green!50!black, solid]table[x=t_s, y=drdT2]{Figures/Bloch_sphere_T1T2_r0m1_pio4.txt};
	
	\node[anchor=north east, fill=white, minimum size=4mm, opacity=0.85, text opacity=1] at (rel axis cs: 1,1) {\textbf{r}};
	\end{axis}
	\end{tikzpicture}
	\fi
	\caption{The spin dynamics in inversion recovery balanced hybrid-state free precession (IR-bHSFP) sequences is depicted on Bloch-spheres (\textbf{a},\textbf{d},\textbf{g},\textbf{j},\textbf{m},\textbf{p}). The polar angle patterns are shown in (\textbf{b},\textbf{e},\textbf{h},\textbf{k},\textbf{n},\textbf{q}), with the color scale providing a reference for the trajectories on the Bloch-sphere. The absolute value of the magnetization (with a negative sign indicating the southern hemisphere) and its normalized derivatives with respect to the relaxation times are the foundation of computing the relative Cram\'er-Rao bound and are shown in (\textbf{c},\textbf{f},\textbf{i},\textbf{l},\textbf{o},\textbf{r}). The optimizations depicted in the left-hand column are limited to $0\leq\vartheta\leq\pi/2$, while the right-hand column show the same optimizations with the limit $0\leq\vartheta\leq\pi/4$.}
	\label{fig:Bloch_sphere}
\end{figure*} 

When considering $T_1$ only, Fig.~\ref{fig:diff_equations}b suggests that there is benefit in keeping the magnetization close to the $z$-axis on the southern hemisphere, and close to the origin on the northern hemisphere, since the absolute value of the derivative is largest in those areas. The derivative with respect to $T_2$, on the other hand, reaches its maximum far away from the $z$-axis (c). Starting from thermal equilibrium, this suggests that it makes sense to acquire signal while the magnetization relaxes along the equator. After the magnetization is brought to zero, the behavior of the derivative induces the magnetization to grow to a comparably large $r$ in order to drive the magnetization back into areas with a large derivative with respect to $T_2$. 

\subsection{Spin Dynamics on the Bloch Sphere}
Fig.~\ref{fig:Bloch_sphere} depicts the hybrid-state spin ensemble trajectories and the corresponding driving functions $\vartheta(t)$ that were numerically optimized for $T_1$ and/or $T_2$ encoding. Some features are shared by all patterns, while others differ depending on the figure of merit. 
In the paragraphs that follow, the common features are described first, followed by the optimization-specific features. 

All patterns are comparatively smooth, which is not explicitly enforced by the optimization. They start with a nonrecurring inversion segment on the southern hemisphere, followed by repetitive loops on the northern hemisphere. The beginning of the first loop differs slightly from that of the other loops, which can be explained by the initial conditions of the loop. At the time when the magnetization crosses the origin (turquoise segments in Fig.~\ref{fig:Bloch_sphere}), the derivatives with respect to the relaxation times are comparatively large. Furthermore, the derivative with respect to $T_2$ has not yet changed its sign. These two properties make this segment a valuable asset to help minimize the correlation between the derivatives with respect to the two relaxation times. As a consequence, the optimized trajectories have a comparatively large polar angle in this segment. 
In contrast, the subsequent loops start with a segment in which the magnetization recovers along the $z$-axis, reflecting small and correlated derivatives. 

Another common feature of all trajectories is that the polar angle decreases rapidly when the shrinking magnetization reaches the steady-state ellipse (magnifications in Fig.~\ref{fig:Bloch_sphere}a,d,g,j,m,p). The arrival at the steady-state ellipse (blue ellipse in Fig.~\ref{fig:Bloch_sphere}) also concludes the last loop in the $T_2$-specific and in the jointly optimized patterns so that they take advantage of the $T_2$-dominated shrinkage of the magnetization (d,g,m,p). By contrast, the $T_1$-specific trajectories (a,j) conclude by maximizing $r$ and $dr/dT_1$ and then briefly bringing the prepared magnetization close to the equator in order to maximize the signal. 

Taking a closer look at the $T_1$-optimized hybrid-state pattern, we observe that the spin trajectory stays close to the $z$-axis on the southern hemisphere and close the origin on the northern hemisphere (Fig.~\ref{fig:Bloch_sphere}a), which is exactly the behavior we expect from intuitive understanding of the derivatives (Section~\ref{sec:Diff_Eq}). 
In fact, the numerical optimization suggests that it is beneficial to forgo any signal at the beginning of the sequence (dark blue segment) in favor of an information-rich signal thereafter. 
Comparing the unconstrained optimization ($0\leq\vartheta\leq\pi/2$) with the constrained one ($0\leq\vartheta\leq\pi/4$), we find that the spin trajectory remains virtually unaffected on the southern hemisphere (Fig.~\ref{fig:Bloch_sphere}j-l). On the northern hemisphere, however, the unconstrained pattern exploits the maximum polar angle of $\pi/2$ to quickly kill the magnetization and create a sort of saturation-recovery loop. When the polar angle is constrained, on the other hand, the magnetization shrinks more slowly and, consequently, fewer loops can be performed in the same amount of time. 

When a hybrid-state sequence is optimized purely for $T_2$ encoding without constraining the polar angle, the magnetization first decays along the equator (Fig.~\ref{fig:Bloch_sphere}d), which is in agreement with the intuitive understanding described in Section~\ref{sec:Diff_Eq}. As $r$ approaches zero, its derivatives with respect to the relaxation times become small as well and the spins follow a trajectory along the $z$-axis (bright blue segment). Thereafter, the spins follows comparatively large loops as predicted. Surprisingly, the optimization does not result in further segments with $\vartheta = \pi/2$. Instead, the trajectory approaches this value only as the magnetization approaches the origin. 
Limiting the polar angle to $0\leq\vartheta\leq\pi/4$ distorts the spin trajectory and the maximum value $\vartheta = \pi/4$ is exploited during several segments (m,n). 

After the inversion pulse, the trajectory of the combined optimization oscillates between large and small $\vartheta$-values. The corresponding derivatives of the magnetization have peaks that are shifted with respect to one another (Fig.~\ref{fig:Bloch_sphere}g-i). 
The size of the loops lies in between the ones of the $T_1$- and $T_2$-optimized trajectories. 
When the polar angle is limited to $0 \leq {\vartheta} \leq \pi/4$, the spin trajectory is distorted in a similar way as for the $T_2$-optimized case.

The hybrid-state equation of spin motion (Eq.~\eqref{eq:r}) can be written in a dimensionless form (Eq.~\eqref{eq:Bloch_r_dimensionless} in the supporting material), which highlights the fact that only the ratios $T_\text{exp} / T_2$ and $T_1 / T_2$ of the involved time constants affect the hybrid-state spin trajectories. 
Varying $T_\text{exp}/T_2$ in the joint optimization results in different trade-offs between the inversion phase and the repetitive loops (see supporting Fig.~\ref{fig:Bloch_sphere_vary_Texp}). Short sequences ($T_\text{exp} \approx T_2$)
require a $T_2$-optimized inversion phase similar to Fig.~\ref{fig:Bloch_sphere}m since the temporal constraints allow only for small, $T_1$-dominated loops.
Increasing the sequence duration allows for larger loops so that the optimized inversion phase increasingly trades $T_2$ for $T_1$ encoding (cf. Fig.~\ref{fig:Bloch_sphere}p). Long sequences ($T_\text{exp} \gg T_2$) use a $T_1$-optimized inversion-recovery segment (cf. Fig.~\ref{fig:Bloch_sphere}j) followed by a combination of large loops that predominantly encode $T_2$ and smaller loops, which provide additional $T_1$ encoding. 

Supporting Fig.~\ref{fig:Bloch_sphere_vary_T1} analyzes the effect of variations of $T_1 / T_2$ on the spin ensemble trajectory and demonstrates that the optimized loop size decreases with an increasing ratio. Large $T_1$ values limit how much the magnetization can grow within a given time. Therefore, the optimized $\vartheta$-pattern spends most of the time for this recovery and the resulting loop size is small. One can further observe that the optimization of $rCRB(T_1) + rCRB(T_2)$ results in different trade-offs between $T_1$ and $T_2$ encoding at different $T_1/T_2$ ratios (Fig.~\ref{fig:Bloch_sphere_vary_T1}c). At intermediate ratios, the algorithm slightly favors the $T_1$-encoding, which is reflected both in the Cram\'er-Rao bound and in the corresponding spin ensemble trajectory, as evident by the inversion-recovery segment. Both at small and at large ratios, the individual Cram\'er-Rao bounds are similar. 

\subsection{The Cram\'er-Rao Bound}
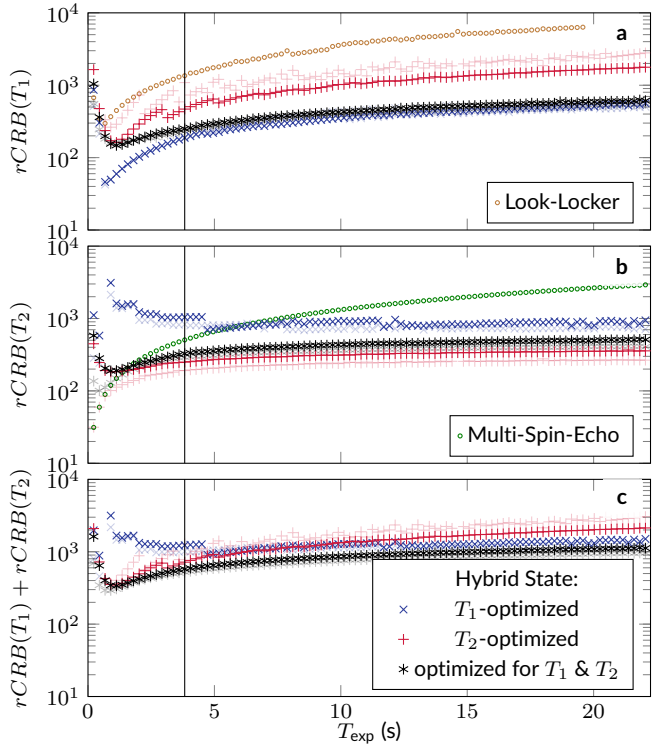
\begin{figure}[tbp]
	\centering
	\ifOL
	\includegraphics[]{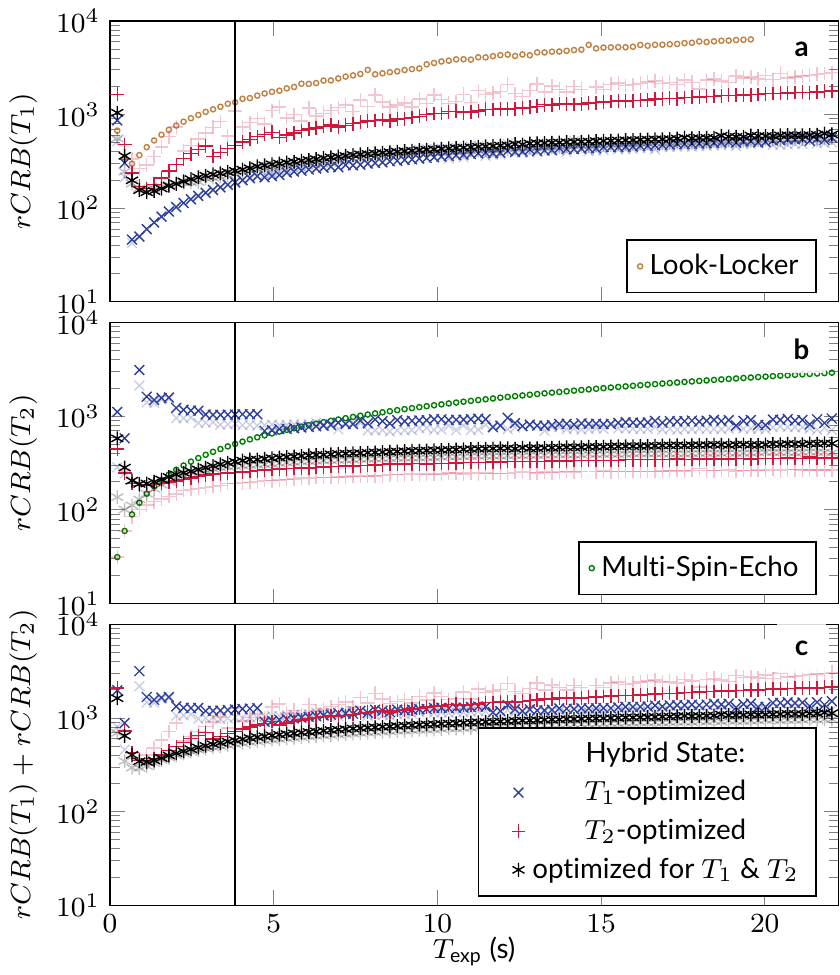}
	\else
	\begin{tikzpicture}[scale = 1]
	%	\small
	\begin{semilogyaxis}[
	width=\columnwidth*0.85,
	height=\textwidth*0.16,
	scale only axis,
	xmin=0,
	xmax=22.25,
	ymin=10,
	ymax=10000,
	xticklabel=\empty,
	ylabel={$rCRB(T_1)$},
	legend entries = {Look-Locker},
	legend pos = south east,
	name=T1,
	clip marker paths=true
	]
	
%	\addplot [color=UKLblue, solid, forget plot, forget plot]table[x=T_s, y=CRB]{Figures/CRB_T1_DESPOT_opt.txt};
%	\addplot [color=brown, mark=o, only marks, mark size=0.75, forget plot]table[x=T_s, y=CRB]{Figures/CRB_T1_IR_bSSFP.txt};
%	
%	\addplot [color=green!50!black, mark=x, only marks, forget plot]table[x=T_s, y=CRB]{Figures/CRB_T1_periodic_pio4.txt};
%	\addplot [color=black, mark=asterisk, only marks, forget plot]table[x=T_s, y=CRB]{Figures/CRB_T1_anti_periodic_pio4.txt};
%	\addplot [color=UKLred, mark=+, only marks, forget plot]table[x=T_s, y=CRB]{Figures/CRB_T1_r0m1_pio4.txt};
%	
%	\addplot [color=purple, densely dashed]table[x=T_s, y=CRB]{Figures/CRB_T1_DESPOT_org.txt};
%	\addplot [color=violet, mark=square*, only marks]table[x=T_s, y=CRB]{Figures/CRB_T1_orgMRF.txt};
%	\addplot [color=cyan, mark=triangle*, only marks]table[x=T_s, y=CRB]{Figures/CRB_T1_orgpSSFP.txt};

\addplot [color=brown, mark=o, only marks, mark size=0.75]table[x=T_s, y=CRB_T1] {Figures/CRB_opt_T1_LookLocker.txt};
	
	\addplot [color=UKLblue, mark=x, only marks, opacity=0.25]table[x=T_s, y=CRB_T1]{Figures/CRB_opt_T1_r0m1_pio2.txt};
	\addplot [color=UKLred, mark=+, only marks, opacity=0.25]table[x=T_s, y=CRB_T1]{Figures/CRB_opt_T2_r0m1_pio2.txt};
	\addplot [color=black, mark=asterisk, only marks, opacity=0.25]table[x=T_s, y=CRB_T1]{Figures/CRB_opt_T1T2_r0m1_pio2.txt};
	
	\addplot [color=UKLblue, mark=x, only marks]table[x=T_s, y=CRB_T1]{Figures/CRB_opt_T1_r0m1_pio4.txt};
	\addplot [color=UKLred, mark=+, only marks]table[x=T_s, y=CRB_T1]{Figures/CRB_opt_T2_r0m1_pio4.txt};
	\addplot [color=black, mark=asterisk, only marks]table[x=T_s, y=CRB_T1]{Figures/CRB_opt_T1T2_r0m1_pio4.txt};
	
	\addplot[color=black, solid, forget plot] coordinates {(3.825, 10) (3.825, 10000)};
	
	\node[minimum size=5mm, opacity=0.85, text opacity=1] at (axis cs:  21.125, 5000) {\textbf{a}};
	\end{semilogyaxis}
	
	\begin{semilogyaxis}[
	width=\columnwidth*0.85,
	height=\textwidth*0.16,
	scale only axis,
	xmin=0,
	xmax=22.25,
	ymin=10,
	ymax=10000,
	xticklabel=\empty,
	ylabel={$rCRB(T_2)$},
	legend entries = {Multi-Spin-Echo},
	legend pos = south east,
	name=T2,
	at=(T1.below south east),
	anchor= north east,
	clip marker paths=true
	]
%	\addplot [color=UKLblue, solid]table[x=T_s, y=CRB]{Figures/CRB_T2_DESPOT_opt.txt};
%	\addplot [color=brown, mark=o, only marks, mark size=0.75]table[x=T_s, y=CRB]{Figures/CRB_T2_IR_bSSFP.txt};
%	
%	\addplot [color=green!50!black, mark=x, only marks, forget plot]table[x=T_s, y=CRB]{Figures/CRB_T2_periodic_pio4.txt};
%	\addplot [color=black, mark=asterisk, only marks, forget plot]table[x=T_s, y=CRB]{Figures/CRB_T2_anti_periodic_pio4.txt};
%	\addplot [color=UKLred, mark=+, only marks, forget plot]table[x=T_s, y=CRB]{Figures/CRB_T2_r0m1_pio4.txt};
%	
%	\addplot [color=purple, densely dashed, forget plot]table[x=T_s, y=CRB]{Figures/CRB_T2_DESPOT_org.txt};
%	\addplot [color=violet, mark=square*, only marks, forget plot]table[x=T_s, y=CRB]{Figures/CRB_T2_orgMRF.txt};
%	\addplot [color=cyan, mark=triangle*, only marks, forget plot]table[x=T_s, y=CRB]{Figures/CRB_T2_orgpSSFP.txt};

	\addplot [color=green!50!black, mark=o, only marks, mark size=0.75]table[x=T_s, y=CRB_T2] {Figures/CRB_opt_T2_TSE.txt};

	\addplot [color=UKLblue, mark=x, only marks, opacity=0.25]table[x=T_s, y=CRB_T2]{Figures/CRB_opt_T1_r0m1_pio2.txt};
	\addplot [color=UKLred, mark=+, only marks, opacity=0.25]table[x=T_s, y=CRB_T2]{Figures/CRB_opt_T2_r0m1_pio2.txt};
	\addplot [color=black, mark=asterisk, only marks, opacity=0.25]table[x=T_s, y=CRB_T2]{Figures/CRB_opt_T1T2_r0m1_pio2.txt};

	\addplot [color=UKLblue, mark=x, only marks]table[x=T_s, y=CRB_T2]{Figures/CRB_opt_T1_r0m1_pio4.txt};
	\addplot [color=UKLred, mark=+, only marks]table[x=T_s, y=CRB_T2]{Figures/CRB_opt_T2_r0m1_pio4.txt};
	\addplot [color=black, mark=asterisk, only marks]table[x=T_s, y=CRB_T2]{Figures/CRB_opt_T1T2_r0m1_pio4.txt};
	
	\addplot[color=black, solid] coordinates {(3.825, 10) (3.825, 10000)};
	
	\node[minimum size=5mm, fill=white, opacity=0.85, text opacity=1] at (axis cs:  21.125, 5250) {\textbf{b}};
	\end{semilogyaxis}
	
	\begin{semilogyaxis}[
	width=\columnwidth*0.85,
	height=\textwidth*0.16,
	scale only axis,
	xmin=0,
	xmax=22.25,
	ymin=10,
	ymax=10000,
	xlabel={$T_\text{exp}~\text{(s)}$},
	xlabel style={yshift=0.15cm},
	ylabel={$rCRB(T_1) + rCRB(T_2)$},
	legend entries = {Hybrid State:, $T_1$-optimized, $T_2$-optimized, optimized for $T_1$ \& $T_2$},
	legend pos = south east,
	name=T1T2,
	at=(T2.below south east),
	anchor= north east,
	clip marker paths=true
	]
%	\addplot [color=UKLblue, solid, forget plot]table[x=T_s, y=CRB]{Figures/CRB_T1T2_DESPOT_opt.txt};
%	\addplot [color=brown, mark=o, only marks, mark size=0.75, forget plot]table[x=T_s, y=CRB] {Figures/CRB_T1T2_IR_bSSFP.txt};
%	
	\addlegendimage{empty legend}
%	\addplot [color=green!50!black, mark=x, only marks]table[x=T_s, y=CRB]{Figures/CRB_T1T2_periodic_pio4.txt};
%	\addplot [color=black, mark=asterisk, only marks]table[x=T_s, y=CRB]{Figures/CRB_T1T2_anti_periodic_pio4.txt};
%	\addplot [color=UKLred, mark=+, only marks]table[x=T_s, y=CRB]{Figures/CRB_T1T2_r0m1_pio4.txt};
%	
%	\addplot [color=UKLred, mark=o, mark size=0.75, only marks]table[x=T_s, y=CRB] {Figures/CRB_T1T2_r0m1_pio4_10s_pause.txt};
%	
%	\addplot [color=purple, densely dashed, forget plot]table[x=T_s, y=CRB]{Figures/CRB_T1T2_DESPOT_org.txt};
%	\addplot [color=violet, mark=square*, only marks, forget plot]table[x=T_s, y=CRB]{Figures/CRB_T1T2_orgMRF.txt};
%	\addplot [color=cyan, mark=triangle*, only marks, forget plot]table[x=T_s, y=CRB]{Figures/CRB_T1T2_orgpSSFP.txt};

%	\addplot [color=cyan, mark=o, only marks, mark size=0.75]table[x=T_s, y=CRB] {Figures/CRB_opt_T1T2_IR_bSSFP.txt};	

	\addplot [color=UKLblue, mark=x, only marks, opacity=0.25, forget plot]table[x=T_s, y=CRB_T1T2] {Figures/CRB_opt_T1_r0m1_pio2.txt};
	\addplot [color=UKLred, mark=+, only marks, opacity=0.25, forget plot]table[x=T_s, y=CRB_T1T2] {Figures/CRB_opt_T2_r0m1_pio2.txt};
	\addplot [color=black, mark=asterisk, only marks, opacity=0.25, forget plot]table[x=T_s, y=CRB_T1T2] {Figures/CRB_opt_T1T2_r0m1_pio2.txt};
	
	\addplot [color=UKLblue, mark=x, only marks]table[x=T_s, y=CRB_T1T2]{Figures/CRB_opt_T1_r0m1_pio4.txt};
	\addplot [color=UKLred, mark=+, only marks]table[x=T_s, y=CRB_T1T2]{Figures/CRB_opt_T2_r0m1_pio4.txt};
	\addplot [color=black, mark=asterisk, only marks]table[x=T_s, y=CRB_T1T2]{Figures/CRB_opt_T1T2_r0m1_pio4.txt};
	
	\addplot[color=black, solid, forget plot] coordinates {(3.825, 10) (3.825, 10000)};
	
	\node[minimum size=5mm, fill=white, opacity=0.85, text opacity=1] at (axis cs:  21.125, 5500) {\textbf{c}};
	\end{semilogyaxis}
	\end{tikzpicture}
	\fi
	\caption{The depicted relative Cram\'er-Rao bounds are defined by Eqs.~\eqref{eq:rCRB_T1} and \eqref{eq:rCRB_T2} and can be understood as a lower bound of the squared inverse SNR efficiency per unit time. They result from numerical optimization for $rCRB(T_1)$, $rCRB(T_2)$, and $rCRB(T_1) + rCRB(T_2)$, while limiting the polar angle to $0 \leq {\vartheta} \leq \pi/2$ (transparent marks) and $0 \leq {\vartheta} \leq \pi/4$ (solid marks), respectively. }
	\label{fig:rCRB_TC}
\end{figure}

This section analyzes the SNR-efficiency of different optimized spin ensemble trajectories. Examining $rCRB(T_1)$, we can observe that the purely $T_1$-optimized trajectories shine most at short experiment durations $T_\text{exp}$ (Fig.~\ref{fig:rCRB_TC}a). As the duration increases, the advantage of the $T_1$-optimized trajectories over the joint optimizations starts to vanish. At $T_\text{exp} = 3.8$s, which is used for the in vivo experiments, we expect only a minor advantage of the $T_1$-optimized sequence in comparison to the joint optimization. In contrast, the $T_2$-optimized sequence shows a poor $T_1$-encoding power. 
Furthermore, we can see that the limit $\vartheta \leq \pi/4$ has only minor effects on the $T_1$ and joint optimizations (transparent vs. solid marks in Fig.~\ref{fig:rCRB_TC}). 
For comparison, we also show the Cram\'er-Rao bound of an optimized Look-Locker sequence \cite{Look1970} in which we also allowed the flip angle to vary over time. One can observe a substantially worse performance compared to the hybrid state, even if one is interested only in $T_1$. 

Examining $rCRB(T_2)$, we also find the difference of the purely $T_2$-optimized and the jointly optimized sequences to be rather small (Fig.~\ref{fig:rCRB_TC}b), especially when limiting the polar angle to $\pi/4$. The $T_1$-optimized patterns, on the other hand, show a poor $T_2$-encoding power. 
For comparison, we also depict the Cram\'er-Rao bound of a multi-spin-echo sequence. For short experiments, this sequence has the same $rCRB(T_2)$ as the hybrid-state sequence that was optimized for $T_2$ without constraining $\vartheta$. In fact, both optimizations result in the same sequences in this limit (not shown here). For long experiments, the performance of the multi-spin-echo sequence does, however, degrade and the more general hybrid state allows for a more efficient sequence design.

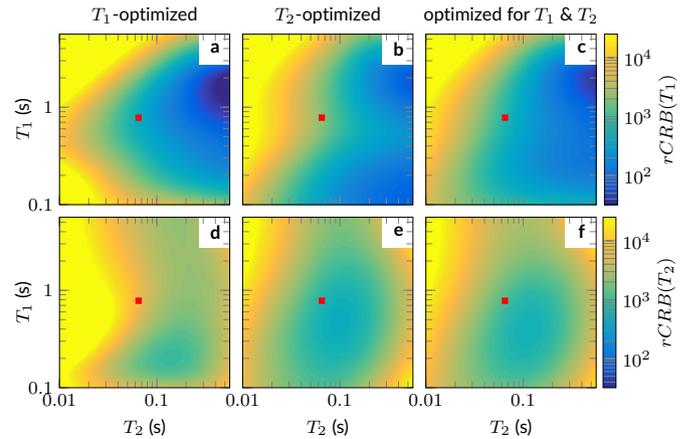
\begin{figure}[bp]
	\centering
	\ifOL
	\includegraphics[]{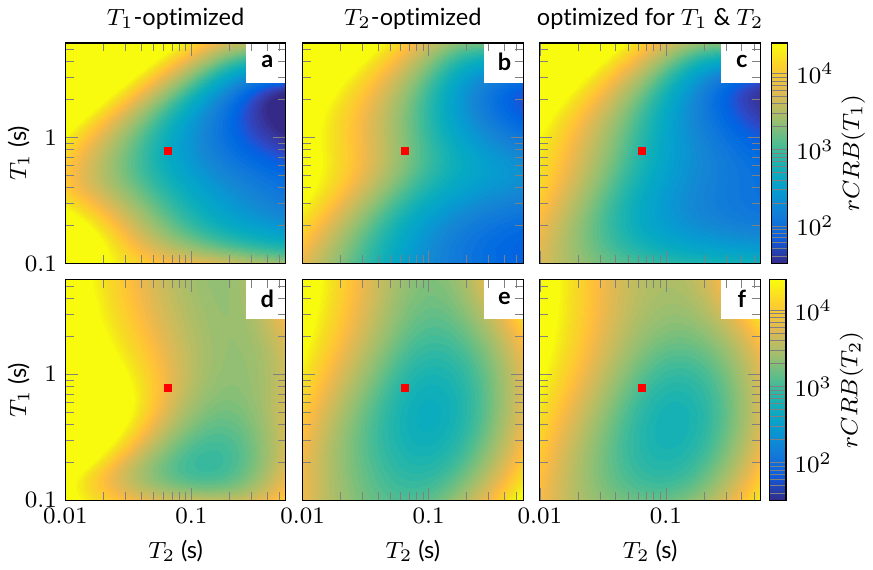}
	\else
	\begin{tikzpicture}[scale=.85]
	\def\s{75}
	
%	\begin{loglogaxis}[
%	width=\s,
%	height=\s,
%	axis on top,
%	scale only axis,
%	log ticks with fixed point,
%	xmin=0.01,
%	xmax={0.5623413252},
%	ymin=0.1,
%	ymax={5.623413252},
%	xticklabels = {},
%	ylabel = {$T_1$~(s)},
%	name = T1_ini,
%	title={original pSSFP},
%	title style={yshift = -0.15cm},
%	]
%	\addplot graphics [xmin=0.01,xmax={0.5623413252},ymin=.1,ymax={5.623413252}] {Figures/rCRB_T1_r0m1_pio4_ini.png};
%	
%	\addplot[mark=square*, only marks, mark options={scale=0.5pt, fill=red, draw=red, thick}] coordinates {(0.065, 0.7810)};
%	\node[inner sep=0mm, minimum size=5mm, fill=white, text=black] at (axis cs:0.4,4) {\textbf{a}};
%	\end{loglogaxis}
	
	\begin{loglogaxis}[
	width=\s,
	height=\s,
	axis on top,
	scale only axis,
	log ticks with fixed point,
	xmin=0.01,
	xmax={0.5623413252},
	ymin=0.1,
	ymax={5.623413252},
	xticklabels = {},
	name = T1_T1opt,
%	at=(T1_ini.north east),
%	anchor=north west,
%	xshift = 0.2cm,
	ylabel = {$T_1$~(s)},
	ylabel style={yshift = -0.25cm},
	title={$T_1$-optimized},
	title style={yshift = -0.15cm},
	]
	\addplot graphics [xmin=0.01,xmax={0.5623413252},ymin=.1,ymax={5.623413252}] {Figures/rCRB_T1_r0m1_pio4_T1opt.png};
	
	\addplot[mark=square*, only marks, mark options={scale=0.5pt, fill=red, draw=red, thick}] coordinates {(0.065, 0.7810)};
	\node[inner sep=0mm, minimum size=5mm, fill=white, text=black] at (axis cs:0.4,4) {\textbf{a}};
	\end{loglogaxis}
	
	\begin{loglogaxis}[
	width=\s,
	height=\s,
	axis on top,
	scale only axis,
	log ticks with fixed point,
	xmin=0.01,
	xmax={0.5623413252},
	ymin=0.1,
	ymax={5.623413252},
	xticklabels = {},
	yticklabels = {},
	name = T1_T2opt,
	at=(T1_T1opt.north east),
	anchor=north west,
	xshift = 0.2cm,
	title={$T_2$-optimized},
	title style={yshift = -0.15cm},
	]
	\addplot graphics [xmin=0.01,xmax={0.5623413252},ymin=.1,ymax={5.623413252}] {Figures/rCRB_T1_r0m1_pio4_T2opt.png};
	
	\addplot[mark=square*, only marks, mark options={scale=0.5pt, fill=red, draw=red, thick}] coordinates {(0.065, 0.7810)};
	\node[inner sep=0mm, minimum size=5mm, fill=white, text=black] at (axis cs:0.4,4) {\textbf{b}};
	\end{loglogaxis}
	
	\begin{loglogaxis}[
	width=\s,
	height=\s,
	axis on top,
	scale only axis,
	log ticks with fixed point,
	xmin=0.01,
	xmax={0.5623413252},
	ymin=0.1,
	ymax={5.623413252},
	xticklabels = {},
	yticklabels = {},
	colormap name = parula,
	colorbar,
	%	colorbar sampled={surf,shader=interp},
	point meta min=32,
	point meta max=25119,
	colorbar style={ymode=log, ylabel=$rCRB (T_1)$, width=0.2cm, xshift=-0.3cm, ylabel=$rCRB (T_1)$, ylabel style={yshift = 0.1cm}},
	name = T1_T1T2opt,
	at=(T1_T2opt.north east),
	anchor=north west,
	xshift = 0.2cm,
	title={optimized for $T_1$ \& $T_2$},
	title style={yshift = -0.15cm},
	]
	\addplot graphics [xmin=0.01,xmax={0.5623413252},ymin=.1,ymax={5.623413252}] {Figures/rCRB_T1_r0m1_pio4_T1T2opt.png};
	
	\addplot[mark=square*, only marks, mark options={scale=0.5pt, fill=red, draw=red, thick}] coordinates {(0.065, 0.7810)};
	\node[inner sep=0mm, minimum size=5mm, fill=white, text=black] at (axis cs:0.4,4) {\textbf{c}};
	\end{loglogaxis}
	
%	\begin{loglogaxis}[
%	width=\s,
%	height=\s,
%	axis on top,
%	scale only axis,
%	log ticks with fixed point,
%	xmin=0.01,
%	xmax={0.5623413252},
%	ymin=0.1,
%	ymax={5.623413252},
%	xlabel = {$T_2$~(s)},
%	ylabel = {$T_1$~(s)},
%	name = T2_ini,
%	at=(T1_ini.below south west),
%	anchor=north west,
%	]
%	\addplot graphics [xmin=0.01,xmax={0.5623413252},ymin=.1,ymax={5.623413252}] {Figures/rCRB_T2_r0m1_pio4_ini.png};
%	
%	\addplot[mark=square*, only marks, mark options={scale=0.5pt, fill=red, draw=red, thick}] coordinates {(0.065, 0.7810)};
%	\node[inner sep=0mm, minimum size=5mm, fill=white, text=black] at (axis cs:0.4,4) {\textbf{e}};
%	\end{loglogaxis}
	
	\begin{loglogaxis}[
	width=\s,
	height=\s,
	axis on top,
	scale only axis,
	log ticks with fixed point,
	xmin=0.01,
	xmax={0.5623413252},
	ymin=0.1,
	ymax={5.623413252},
	xlabel = {$T_2$~(s)},
	ylabel = {$T_1$~(s)},
	ylabel style={yshift = -0.25cm},
	name = T2_T1opt,
	at=(T1_T1opt.below south west),
	anchor=north west,
	]
	\addplot graphics [xmin=0.01,xmax={0.5623413252},ymin=.1,ymax={5.623413252}] {Figures/rCRB_T2_r0m1_pio4_T1opt.png};
	
	\addplot[mark=square*, only marks, mark options={scale=0.5pt, fill=red, draw=red, thick}] coordinates {(0.065, 0.7810)};
	\node[inner sep=0mm, minimum size=5mm, fill=white, text=black] at (axis cs:0.4,4) {\textbf{d}};
	\end{loglogaxis}
	
	\begin{loglogaxis}[
	width=\s,
	height=\s,
	axis on top,
	scale only axis,
	log ticks with fixed point,
	xmin=0.01,
	xmax={0.5623413252},
	ymin=0.1,
	ymax={5.623413252},
	xlabel = {$T_2$~(s)},
	yticklabels = {},
	name = T2_T2opt,
	at=(T2_T1opt.north east),
	anchor=north west,
	xshift = 0.2cm,
	]
	\addplot graphics [xmin=0.01,xmax={0.5623413252},ymin=.1,ymax={5.623413252}] {Figures/rCRB_T2_r0m1_pio4_T2opt.png};
	
	\addplot[mark=square*, only marks, mark options={scale=0.5pt, fill=red, draw=red, thick}] coordinates {(0.065, 0.7810)};
	\node[inner sep=0mm, minimum size=5mm, fill=white, text=black] at (axis cs:0.4,4) {\textbf{e}};
	\end{loglogaxis}
	
	\begin{loglogaxis}[%
	width=\s,
	height=\s,
	axis on top,
	scale only axis,
	log ticks with fixed point,
	xmin=0.01,
	xmax={0.5623413252},
	ymin=0.1,
	ymax={5.623413252},
	xlabel = {$T_2$~(s)},
	yticklabels = {},
	colormap name = parula,
	colorbar,
	%	colorbar sampled={surf,shader=interp},
	point meta min=32,
	point meta max=25119,
	colorbar style={ymode=log, ylabel=$rCRB (T_2)$, width=0.2cm, xshift=-0.2cm, ylabel=$rCRB (T_2)$, ylabel style={yshift = 0.1cm}},
	name = T2_T1T2opt,
	at=(T2_T2opt.north east),
	anchor=north west,
	xshift = 0.2cm,
	]
	\addplot graphics [xmin=0.01,xmax={0.5623413252},ymin=.1,ymax={5.623413252}] {Figures/rCRB_T2_r0m1_pio4_T1T2opt.png};
	
	\addplot[mark=square*, only marks, mark options={scale=0.5pt, fill=red, draw=red, thick}] coordinates {(0.065, 0.7810)};
	\node[inner sep=0mm, minimum size=5mm, fill=white, text=black] at (axis cs:0.4,4) {\textbf{f}};
	\end{loglogaxis}
	\end{tikzpicture}
	\fi
	\caption{The performance of IR-bHSFP sequences optimized with $0 \leq \vartheta \leq \pi/4$ (Fig.~\ref{fig:Bloch_sphere}j-r) is illustrated through plots of the relative Cram\'er-Rao bounds, which provide a lower bound for the noise in the estimated relaxation times. All patterns were optimized for $T_1 = 781~\text{ms}$ and $T_2 = 65~\text{ms}$, as indicated by the red square, and were tested for the entire parameter space in a sample MRF dictionary. Note the logarithmic scale in all three dimensions.}
	\label{fig:rCRB_T1_T2_IR}
\end{figure}

So far, the Cram\'er-Rao bound was analyzed only at the specific $T_1 = 781~\text{ms}$ and $T_2 = 65~\text{ms}$ that were used for the optimization. Fig.~\ref{fig:rCRB_T1_T2_IR} analyzes the performance of sequences optimized for these particular relaxation times over a larger parameter space, i.e. over the range of $T_1$ and $T_2$ values commonly found in biological tissue.  
We found that optimizing for a single set of relaxation times does not merely result in good performance in the proximity of this set of parameters, which is visualized by the red dot. Instead, we found good performance over a large area in the $T_1$-$T_2$-space with the lowest Cram\'er-Rao bound not being located at the relaxation times used for the optimizations.

In the supplementary material, we bridge the gap between the Cram\'er-Rao bound and the correlation coefficients, which are used to reconstruct the parameter maps in MRF \cite{Ma2013,Asslander2018} (Fig.~\ref{fig:correlation_WM} and \ref{fig:correlation_max}).

\subsection{In Vivo Experiments}
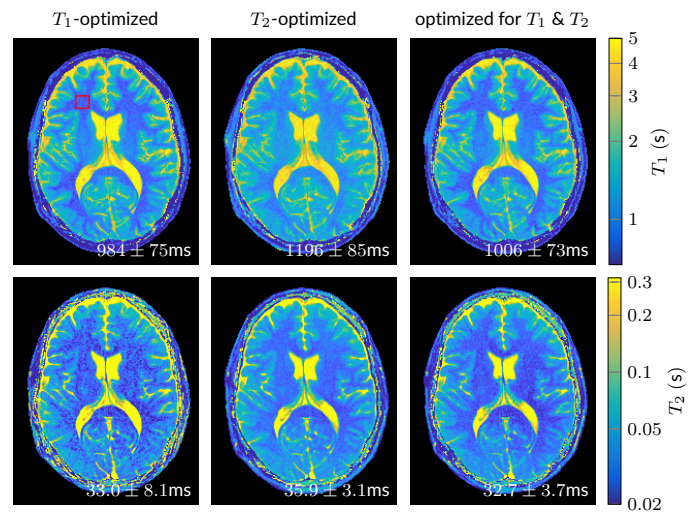
\begin{figure}[tb]
	\centering
	\ifOL
	\includegraphics[]{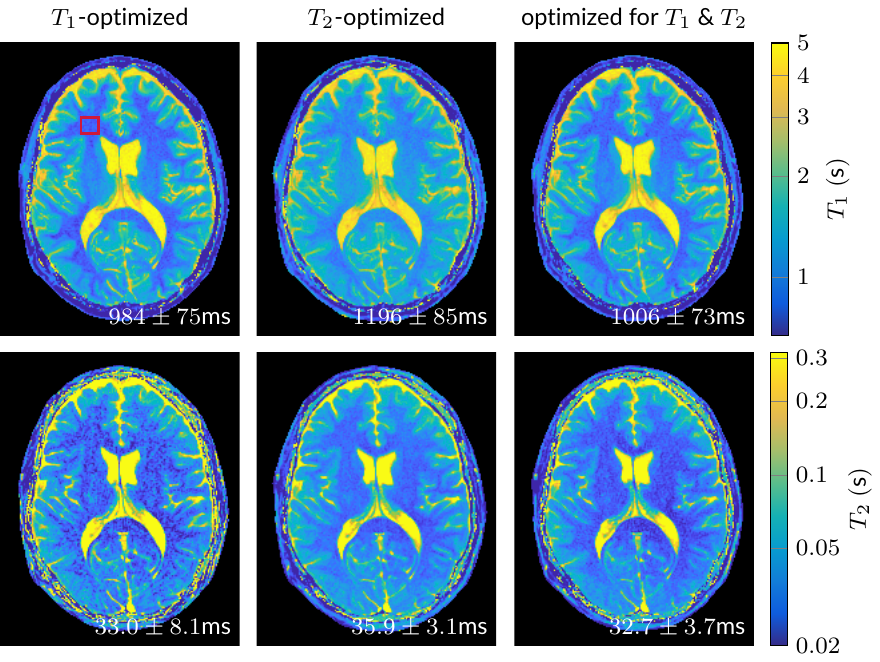}
	\else
	\begin{tikzpicture}[scale=.85]
	\def\s{3.5cm}

	\begin{axis}[%
	width={\s/220*180},
	height=\s,
	axis on top,
	scale only axis,
	xmin=0,
	xmax=1,
	ymin=0,
	ymax=1,
	hide axis,
	name=T1_T1opt,
%	at=(T1_pSSFP.right of north east),
%	anchor= left of north west,
%	xshift=.08in,
	%	at=(PD_T1opt.below south west),
	%	anchor=above north west,
	%	yshift=-.08in,
	title={$T_1$-optimized},
	title style={yshift = -0.15cm},
	]
	\addplot graphics [xmin=0,xmax=1,ymin=0,ymax=1] {Figures/T1_9100-1.png};
	\draw [UKLred, thick](axis cs: {61/180}, {164/220}) rectangle (axis cs: {74/180}, {152/220});
	\node[text=white, anchor = south east]at (axis cs:1, 0) {$984 \pm 75$ms};
	\end{axis}
	
	\begin{axis}[%
	width={\s/220*180},
	height=\s,
	axis on top,
	scale only axis,
	xmin=0,
	xmax=1,
	ymin=0,
	ymax=1,
	hide axis,
	name=T2_T1opt,
	at=(T1_T1opt.below south west),
	anchor=above north west,
	yshift=-.08in,
	]
	\addplot graphics [xmin=0,xmax=1,ymin=0,ymax=1] {Figures/T2_9100-1.png};
	\node[text=white, anchor = south east]at (axis cs:1, 0) {$33.0 \pm 8.1$ms};
	\end{axis}
	
	%	\begin{axis}[%
	%	width={\s/220*180},
	%	height=\s,
	%	axis on top,
	%	scale only axis,
	%	xmin=0,
	%	xmax=1,
	%	y dir=reverse,
	%	ymin=0,
	%	ymax=1,
	%	hide axis,
	%	name=PD_T2opt,
	%	at=(PD_T1opt.right of north east),
	%	anchor= left of north west,
	%	xshift=.08in,
	%	point meta min=0.01,
	%	point meta max=1,
	%	title={$T_2$-optimized},
	%	title style={yshift = -.2cm},
	%	]
	%	\addplot graphics [xmin=0,xmax=1,ymin=0,ymax=1] {Figures/PD_9108-1.png};
	%	\end{axis}
	
	\begin{axis}[%
	width={\s/220*180},
	height=\s,
	axis on top,
	scale only axis,
	xmin=0,
	xmax=1,
	ymin=0,
	ymax=1,
	hide axis,
	name=T1_T2opt,
	at=(T1_T1opt.right of north east),
	anchor= left of north west,
	xshift=.08in,
	%	at=(PD_T2opt.below south west),
	%	anchor=above north west,
	%	yshift=-.08in,
	title={$T_2$-optimized},
	title style={yshift = -0.15cm},
	]
	\addplot graphics [xmin=0,xmax=1,ymin=0,ymax=1] {Figures/T1_9108-1.png};
	\node[text=white, anchor = south east]at (axis cs:1, 0) {$1196 \pm 85$ms};
	\end{axis}
	
	\begin{axis}[%
	width={\s/220*180},
	height=\s,
	axis on top,
	scale only axis,
	xmin=0,
	xmax=1,
	ymin=0,
	ymax=1,
	hide axis,
	name=T2_T2opt,
	point meta min=0,
	point meta max=.25,
	at=(T1_T2opt.below south west),
	anchor=above north west,
	yshift=-.08in,
	]
	\addplot graphics [xmin=0,xmax=1,ymin=0,ymax=1] {Figures/T2_9108-1.png};
	\node[text=white, anchor = south east]at (axis cs:1, 0) {$35.9 \pm 3.1$ms};
	\end{axis}
	
	%	\begin{axis}[%
	%	width={\s/220*180},
	%	height=\s,
	%	axis on top,
	%	scale only axis,
	%	xmin=0,
	%	xmax=1,
	%	y dir=reverse,
	%	ymin=0,
	%	ymax=1,
	%	hide axis,
	%	name=PD_T1T2opt,
	%	at=(PD_T2opt.right of north east),
	%	anchor= left of north west,
	%	xshift=.08in,
	%	colormap/blackwhite,
	%	colorbar,
	%	point meta min=0.01,
	%	point meta max=1,	
	%	colorbar style={ylabel=$PD~(\text{a.u.})$, width=0.5cm, xshift=-0.1cm},
	%	title={optimized for $T_1$ \& $T_2$},
	%	title style={yshift = -.2cm},
	%	]
	%	\addplot graphics [xmin=0,xmax=1,ymin=0,ymax=1] {Figures/PD_9104-1.png};
	%	\end{axis}
	
	\begin{axis}[%
	width={\s/220*180},
	height=\s,
	axis on top,
	scale only axis,
	xmin=0,
	xmax=1,
	ymin=0,
	ymax=1,
	hide axis,
	name=T1_T1T2opt,
	colormap name = parula,
	colorbar,
	point meta min=0.668,
	point meta max=5,
	colorbar style={ymode=log, ytick={1,2,3,4,5}, ylabel=$T_1~(\text{s})$, log ticks with fixed point, ylabel style = {yshift = 0cm}, width=0.2cm, xshift=-0.1cm},
%	point meta min=-.175,
%	point meta max=.7,
%	colorbar style={ytick={-1.3, -1, -0.699, -0.523, -0.301, 0, 0.176, 0.3010, 0.477, 0.699, 1, 1.3010}, yticklabel=${\pgfmathparse{10^\tick}\pgfmathprintnumber\pgfmathresult}$, yticklabel style={/pgf/number format/fixed, /pgf/number format/precision=2}, ylabel=$T_1~(\text{s})$, width=0.2cm, xshift=-0.1cm},
	title={optimized for $T_1$ \& $T_2$},
	title style={yshift = -0.15cm},
	at=(T1_T2opt.right of north east),
	anchor= left of north west,
	xshift=.08in,
	]
	\addplot graphics [xmin=0,xmax=1,ymin=0,ymax=1] {Figures/T1_9104-1.png};
	\node[text=white, anchor = south east]at (axis cs:1, 0) {$1006 \pm 73$ms};
	\end{axis}
	
	\begin{axis}[%
	width={\s/220*180},
	height=\s,
	axis on top,
	scale only axis,
	xmin=0,
	xmax=1,
	ymin=0,
	ymax=1,
	hide axis,
	name=T2_T1T2opt,
	colormap name = parula,
	colorbar,
	point meta min=0.02,
	point meta max=0.316,
	colorbar style={ymode=log, ytick={0.02,0.05,0.1,0.2,0.3}, ylabel=$T_2~(\text{s})$, log ticks with fixed point, ylabel style = {yshift = 0.1cm}, width=0.2cm, xshift=-0.1cm},
%	point meta min=-1.7,
%	point meta max= -0.5,	
%	colorbar style={ytick={-1.699, -1.3, -1.0969, -1, -0.8239, -0.699, -0.523, -0.3979, -0.301, 0, 0.3010, 0.699, 1, 1.3010}, yticklabel=${\pgfmathparse{10^\tick}\pgfmathprintnumber\pgfmathresult}$, yticklabel style={/pgf/number format/fixed, /pgf/number format/precision=2}, ylabel=$T_2~(\text{s})$, width=0.2cm, xshift=-0.1cm},
	at=(T1_T1T2opt.below south west),
	anchor=above north west,
	yshift=-.08in,
	]
	\addplot graphics [xmin=0,xmax=1,ymin=0,ymax=1] {Figures/T2_9104-1.png};
	\node[text=white, anchor = south east]at (axis cs:1, 0) {$32.7 \pm 3.7$ms};
	\end{axis}
	\end{tikzpicture}
	\fi
	\caption{The in vivo data were acquired with the excitation patterns depicted in Fig.~\ref{fig:Bloch_sphere}j-r with the limit $0 \leq \vartheta \leq \pi/4$. The parameter maps have an in-plane resolution of $1~\text{mm}$ and were acquired in $3.8\text{s}$. Note the logarithmic scale of the $T_1$ and $T_2$ color maps. The red rectangle indicates the region of interest used for calculating the mean and standard deviation denoted in the corner of the images.}
	\label{fig:InVivo}
\end{figure}

Fig.~\ref{fig:InVivo} depicts quantitative in vivo maps acquired with the patterns shown in Fig.~\ref{fig:Bloch_sphere}k,n,q and with one radial k-space spoke acquired for each time frame. The total scan time was approximately $3.8$s for each of the three experiments. 
Systematic deviations can be noted in the measured relaxation times (Fig.~\ref{fig:InVivo}). These variations are most likely caused by magnetization transfer \cite{Hilbert2016}, which is not addressed in the present work. 
Focusing on the noise properties, one can observe a good agreement between the Cram\'er-Rao bound predictions and the resulting noise level in the parameter maps. As anticipated, the purely $T_1$-optimized pattern and the jointly-optimized pattern achieve a very similar SNR level in the $T_1$-maps (cf. Fig.~\ref{fig:rCRB_TC}a), while the $T_2$-optimized pattern is slightly worse. The gap is substantially larger in the $T_2$-maps, where the $T_2$-optimized and the jointly-optimized patterns achieve similar performance, but the $T_1$-optimized pattern has a substantially larger standard deviation (cf. the ROI analysis displayed in the corner of Fig.~\ref{fig:InVivo}). 

\section{Discussion} \label{sec:Discussion}
The purpose of this paper is to shed some light on the optimal encoding of spin relaxation times. In hybrid state, the spin dynamics is trapped in a single dimension and can be described by a single, uncoupled differential equation (Eq.~\eqref{eq:r}), which may be conveniently visualized (Fig.~\ref{fig:diff_equations}). Insights gleaned from intuitive understanding of the governing equation were found to align well with numerical optimizations (Fig.~\ref{fig:Bloch_sphere}). 
Numerically searching for the optimal spin trajectories is a non-convex problem so we can only speculate about the optimality of our results. However, the notable correspondence to the intuitive understanding of the governing differential equation, together with the number of reproducible features in the optimized trajectories, gives us confidence that the obtained numerical results are closely related to the truly optimal trajectories. 

The Cram\'er-Rao bound was used as figure of merit since it provides a universal lower bound for the noise transfer from a time series to the estimated parameters. It also provides the flexibility to tailor the acquisition to, for example, encode $T_1$ while accounting for correlations in $T_2$-space. 
Within the limits of this analysis---most notably our neglect of the incomplete spatial encoding and the non-convexity of the optimization problems---we showed that the hybrid state has superior SNR properties compared to traditional methods, namely Look-Locker for $T_1$-encoding and multi-spin-echo sequences for $T_2$-encoding. Moreover, we found that the joint optimizations have nearly the same performance in encoding each relaxation time compared to hybrid-state sequences that are specialized to encode a single parameter. Thus, the second relaxation time can be measured at virtually no extra cost. 

Here, we optimized the sequences only for a single set of relaxation times, which correspond to white matter. However, we found that such optimized sequences perform well over a large range of relaxation times. If one wanted to extend this area further, one could optimize for the sum of the Cram\'er-Rao bounds at different relaxation times. 

Our least-constrained optimizations used the limits $-1 \leq r(t) \leq 1$ and $0 \leq {\vartheta}(t) \leq \pi/2$. This unusual definition of spherical coordinates was chosen to deliberately prohibit magnetization from revisiting the southern hemisphere after passing through the origin. This allows us to implement the RF-pulses without major violations of the small tip-angle approximation. 
However, we demonstrated the benefits of the southern hemisphere and for long experiments as might be required, e.g., for 3D imaging, it might be desirable to revisit the southern hemisphere. This could be achieved by simply changing the limits to $0 \leq r(t) \leq 1$ and $0 \leq {\vartheta}(t) \leq \pi$. However, this would change the optimization landscape, and the convergence behavior might be impaired. Alternatively, one could use anti-periodic boundary conditions, as demonstrated in Ref. \cite{Asslander2018c}. 

The present work focuses on signal-to-noise properties and the employed model is derived from the standard Bloch equation. The in vivo data shown in Fig.~\ref{fig:InVivo} are corrupted by systematic errors such as $B_0$- and $B_1$-inhomogeneities, as well as magnetization transfer effects. The first two problems can be corrected during the reconstruction process \cite{Ma2013,Buonincontri2016,Cloos2016,Ma2017}, and it may be possible to also address magnetization transfer effects using a multi pool model \cite{Hilbert2016}. Future work will include numerical optimizations that incorporate these effects in order to minimize correlations between the corresponding parameters and the spin relaxation times. 

%TODO: conclusion?

%\newpage
%\section*{Acknowledgments}
%This work was supported by the research grants NIH/NIBIB R21 EB020096 and NIH/NIAMS R01 AR070297, and was performed under the rubric of the Center for Advanced Imaging Innovation and Research (CAI2R, www.cai2r.net), a NIBIB Biomedical Technology Resource Center (NIH P41 EB017183).

\bibliography{library}%
\end{document}